# Fluids mobilization in Arabia Terra, Mars: depth of pressurized reservoir from mounds self-similar clustering

Riccardo Pozzobon[1], Francesco Mazzarini[2], Matteo Massironi[1], Angelo Pio Rossi[3], Monica Pondrelli[4], Gabriele Cremonese[5], Lucia Marinangeli[6]

1 Department of Geosciences, Università degli Studi di Padova, Via Gradenigo 6 - 35131, Padova, Italy
2 Istituto Nazionale di Geofisica e Vulcanologia (INGV), Via Della Faggiola, 32 - 56100 Pisa, Italy
3 Department of Physics and Earth Sciences, Jacobs University Bremen, Campus Ring 1 -28759 Bremen, Germany
4 International Research School of Planetary Sciences, Università d'Annunzio, viale Pindaro 42 – 65127, Pescara, Italy
5 INAF, Osservatorio Astronomico di Padova, Vicolo dell'Osservatorio 5 - 35122, Padova, Italy
6 Laboratorio di Telerilevamento e Planetologia, DISPUTer, Universita' G. d'Annunzio, Via Vestini 31 - 66013 Chieti, Italy

Abstract

Arabia Terra is a region of Mars where signs of past-water occurrence are recorded in several landforms. Broad and local scale geomorphological, compositional and hydrological analyses point towards pervasive fluid circulation through time. In this work we focus on mound fields located in the interior of three casters larger than 40 km (Firsoff, Kotido and unnamed crater 20 km to the east) and showing strong morphological and textural resemblance to terrestrial mud volcanoes and spring-related features. We infer that these landforms likely testify the presence of a pressurized fluid reservoir at depth and past fluid upwelling. We have performed morphometric analyses to characterize the mound morphologies and consequently retrieve an accurate automated mapping of the mounds within the craters for spatial distribution and fractal clustering analysis. The outcome of the fractal clustering yields information about the possible extent of the percolating fracture network at depth below the craters. We have been able to constrain the depth of the pressurized fluid reservoir between ~2.5 and 3.2 km of depth and hence, we propose that mounds and mounds alignments are most likely associated to the presence of fissure ridges and fluid outflow. Their process of formation is genetically linked to the formation of large intra-crater bulges previously interpreted as large scale spring deposits. The overburden removal caused by the impact crater formation is the inferred triggering mechanism for fluid pressurization and upwelling, that through time led to the formation of the intra-crater **bulges and, after compaction and sealing, to the widespread mound fields in their surroundings.**

1 Introduction

Mud volcanoes on Earth, such as Dashgil mud volcanoes in Azerbaijan (Bonini, 2012; Skinner and Mazzini, 2009) form when deep thick sedimentary sequences undergo high pore fluid pressures often triggered by compaction through loading or tectonic deformation. The mobilized fluids often tend to exploit pre-existing fracture networks (e.g. Dimitrov et al., 2002; Kopf, 2002; Skinner and Mazzini, 2009; Bonini, 2012; Bonini and Mazzarini, 2010; Oehler and Etiope, 2017). They are inferred to be the surface expression of deeply rooted vertical structures where sediment extrusion is driven by a mobile fluid fraction (e.g., water, hydrocarbons, gas), migrating upward from reservoirs several kilometers deep (Deville et al., 2003). Fluid expulsion features may involve clastic (i.e., mud volcanism) or evaporitic depositional processes related to the rise of oversaturated fluids (i.e., spring mounds).

Putative mud volcanic fields on Mars were spotted in several areas (e.g., Arabia Terra, Acidalia Planitia, Isidis Planitia, Utopia Basin, Chryse Planitia and Galaxias Phossae) presenting both pitted cones and smooth mound morphologies (Skinner and Mazzini, 2009; Oehler and Allen, 2010; Komatsu et al., 2015; Okubo, 2016).

The favored hypothesis for the formation of large mud volcanic fields, such as Acidalida and Chryse Planitia (that presents more than 18.000 pitted cones), is the interplay between high rates of sediment deposition due to outflow channel activity and subsequent compaction (Tanaka et al., 2013a). High pressure is created at depth and leads to disruption of the sedimentary sequence along fractures that act as conduits for expulsion of mud breccia (Skinner and Mazzini 2009, Bonini, 2012, Okubo 2016, Oehler and Etiope, 2017). Skinner and Mazzini (2009) show that in some cases pre-existing structural grain in the Martian crust can control occurrence of mounds alignments.

In the presence of thick sedimentary sequences with a deep source associated with long-term compaction, fluid extrusion rates control the cones' morphology; high extrusion rates form pitted cones, whereas low extrusion rates generate un-pitted domes (Skinner and Mazzini, 2009; Bonini 2012; Allen and Oehler, 2008; Okubo 2016; Pondrelli et al., 2011; Oehler and Etiope, 2017).

Impact craters can also play an important role in the generation of mud volcanoes and springs since they strongly modify the planetary landscape and surface/subsurface hydrology (e.g. Skinner and

Mazzini, 2009; Rodríguez et al., 2005, Carrozzo et al., 2017). It has been demonstrated that impact processes produce a pervasive network of fractures (Melosh, 2007; Collins et al., 2004-2011; Wunneman et al., 2006) that increase the secondary permeability and thus favor fluid circulation within the crust (Oehler and Etiope, 2017). Moreover, it has been hypothesized the presence of long-lasting impact-induced hydrothermalism on Mars and its duration was calculated for several crater diameters (Abramov and King, 2005) that can last up to 380,000 years in ~200 km-wide basins and was shown in the case of Auki crater in Carrozzo et al., (2017). Terrestrial mud volcanoes indeed require a complex plumbing system of interconnected fractures that enable liquids to flow and also the presence of gas driven eruption phases (mostly hydrocarbons and $CO_2$). Several authors have detected methane on Martian atmosphere interpreted as episodically produced from unknown sources (Mumma et al., 2009; Geminale et al., 2008; Formisano et al., 2005). Oze et al., (2005) proposed that one possible source of methane can be related to the hydrothermal activity caused by latent heat localized in impact craters. This phenomenon could produce methane and $CO_2$ as a consequence of the hydrothermal alteration of basalts. Moreover, the presence of ancient super-volcanoes in northern Arabia Terra could have provided sufficient heat and fluids to trigger mud volcanic activity (Michalski et al., 2013). In all cases, the actual presence and depth of the fluid reservoir feeding the mounds within Arabia craters is heavily debated as well as the interpretation for the sources of methane degassing (Webster et al., 2015 and references therein). Nonetheless in the framework of ExoMars – Trace Gas Orbiter mission – locating and characterizing the potential outgassing sites is of primary importance. For this reason, we approached the study of mound fields in the Firsoff, Kotido, and the easternmost unnamed crater with different techniques including surface analysis on DTMs and high-resolution images, and investigation of possible plumbing systems though self-similar clustering and fractal analysis. However, the correct interpretation of mound features as mud volcanoes is particularly biased where image resolution is not sufficient to distinguish fine details and textures; thus, we have provided a well constrained mapping rationale of mud volcanic morphologies though their morphometric characterization based on the numerical characterization of the mounds already interpreted as mud volcanoes on high resolution images (i.e., HiRISE) (Pondrelli et al., 2011, Pondrelli et al., 2015). Extrapolating the obtained morphometric parameters to neighboring craters we were able to extract all similar mounds (i.e., potential mud-volcanoes) whose distribution was afterwards analyzed through self-similar clustering to derive the depth of the related fluid source. The effectiveness of the self-similar (fractal) clustering approach has been proved both on Earth on monogenic volcanic vents along the East African Rift as well as on Ascraeus Mons dykes in Mars (Mazzarini and Isola, 2010; Mazzarini et al., 2013; Pozzobon et al., 2014). Indeed, monogenic vents and dykes show a self-similar clustering (with a fractal exponent) in a defined size range comprised between a lower and upper cutoff, the value of the upper cutoff closely matches the actual depth of the magmatic reservoir (i.e. Mazzarini and Isola, 2010). This approach has been successfully applied to derive the depth of pressurized layers feeding mud-volcanoes in the foreland of the Greater Caucasus in Azerbaijan (Bonini and Mazzarini, 2010) and on a wide range of mound-like morphologies on Martian surface (i.e., pingos, rootless cones, monogenic magmatic cones, mud volcanoes; De Toffoli et al., 2018). Overall, our results will provide a better understanding of the hydrological cycle, the formation, and most of all, the factors driving these processes (as in Zabrusky et al., 2013, Andrews Hanna et al., 2011). It will also provide clues to why these are so spatially localized, a topic that is still matter of considerable debate.

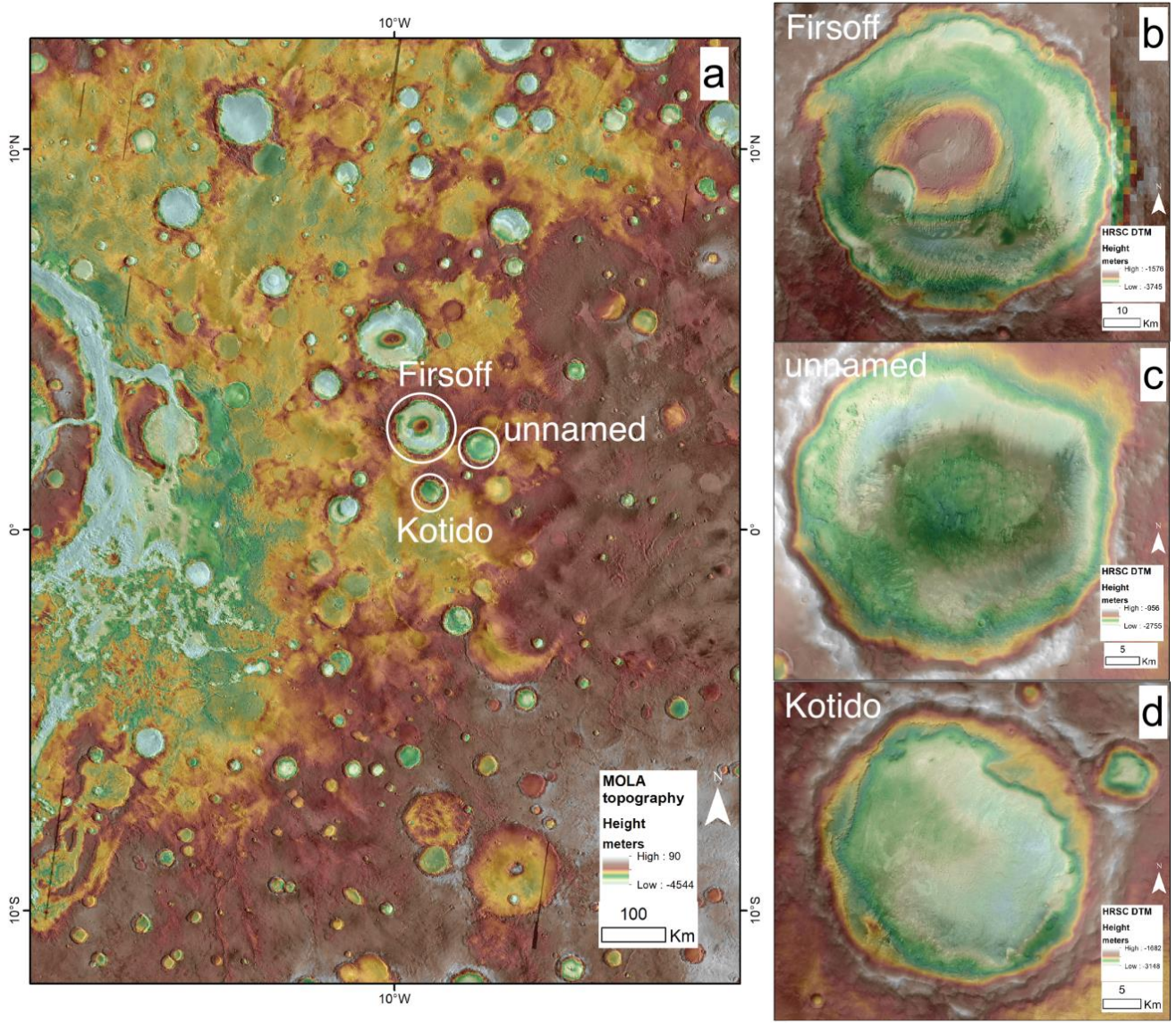

**Fig.1**: Arabia Terra with THEMIS IR Day and MOLA MEGDR topography in transparency. The three studied craters are highlighted with white circles. The right section of the panel shows Firsoff, Kotido, and the eastern unnamed crater with HRSC nadir image and HRSC DTM in transparency. Note the central bulge in all of the craters.

## 2 Geologic framework

*2.1 Regional setting*

Arabia Terra region (~3000 km$^2$) extends from the southern heavily cratered highlands to the northern lowlands, gently dipping (0.09°) northward with an elevation drop of 4 km over a distance of 2500 km (Fig. 1). Within several impact craters in the studied area (0°25'N to 3°25'N and 7°07'W to 10°27'W) numerous small mounds, pitted cones and knobs are present (Pondrelli et al., 2011, 2015, Allen and Oehler, 2008) as well as kilometer-size features interpreted as spring mounds (Allen and Oehler, 2009). All these features are embedded within light albedo large intra-crater bulges (up to tens of kilometers in diameter) with thin quasi-periodic layering in some cases (Lewis and Aharonson, 2014). Overall on Mars, nearly 60% of bulged craters are located in Arabia Terra (Bennett and Bell, 2016). Several hypotheses have been proposed for the origin of intra-crater bulges, such as large-scale spring deposits (Rossi et al., 2008), aeolian infilling due to katabatic winds (Kite et al., 2013), and strong wind erosion of sedimentary infilling (Bennett and Bell, 2016, Zabrusky et al., 2012, Andrews-Hanna et al., 2011) and transition in the depositional and subsequent erosion from dry to wet conditions (Kite et al., 2016).

All the craters within this Arabia Terra sector are embedded within the so-called Cratered Unit (CU, Tanaka et al., 2014; Pondrelli et al., 2015), a Noachian plateau sequence consisting of pyroclastics, lava flows and brecciated material (Scott and Tanaka, 1986).

The deposits found within crater interiors can be classified into two major units: a layered unit

(ELD) and a mounds unit (MU). The ELD consists of a high albedo layered material, in some cases interbedded with darker material often disrupted in a polygonal pattern that overprints the original deposit, sometimes resembling the etched terrain seen in Meridiani Planum (Hynek et al., 2002). This entire unit has gentle dip angles that appear to be adapted to the pre-existing topography and mantling the inner crater terrace. The layered sequences can reach up to 2 km of thickness inside craters (measured within Firsoff) whereas they are much less pronounced in the outer plateau. Here, the sedimentary succession reaches a maximum thickness of 10 m and unconformably overlies the CU (Pondrelli et al., 2015, Franchi et al., 2014). In places, the ELD unit is buried by a Hummocky Material unit likely made of volcanic dark-toned rocks (Franchi et al., 2014, Pondrelli et al., 2011, 2015). In the southern sector of Arabia Terra, Hesperian flood basalts (Ridged Plain Materials Unit) bury entirely all the previously described successions (Scott and Tanaka, 1986). Therefore, it appears clear that the ELDs are stratigraphically constrained between the Noachian plateau sequence and the Hesperian flood basalts in a time range when liquid water was stable on Martian surface and subsurface, creating lacustrine and fluvial landforms as well as large amounts of alteration minerals (e.g. Flahaut et al., 2015, Pondrelli et al., 2015, Franchi et al., 2014).

The MU unit is associated with, and found within ELDs and presents mounds of few hundred meters in diameter, consisting of a layered/non-layered breccia mixed with fine grain matrix (Pondrelli et al., 2011) and sharing similar compositional characters with ELD (Pondrelli et al., 2015).

*2.2 Case study craters*

We concentrated our efforts on characterizing and mapping mounds in three adjacent craters with a widespread presence of ELDs and mounds.

The Firsoff crater (2.61°N–9.21°E) has a diameter of ~90 km and its interior contains widespread ELDs and a large central bulge 35-40 km in diameter and ~2 km high. The crater's interior is formed essentially of ELD and MU units and appears to be strongly affected by degradation, showing a number of erosional features and aeolian deposits such as dark sand deposits. Most of the mounds are concentrated in the south-eastern inner floor of the crater, where they appear fresher and less degraded by erosion **than** in the eastern sector where they are less numerous (Pondrelli et al., 2015; Pondrelli et al., 2011). In the other sectors of the crater, mounds are still present but a clear visual (distinction) separation/categorization based on erosive morphologies is difficult.

**The** Kotido crater (11°N–91°E) is located **to the South of the** Firsoff crater and has a diameter of ~40 km with a depth between 800 and 1000 m below the regional surface. A well-preserved ELD formation covers its floor (Franchi et al., 2014; Pondrelli et al., 2015). HRSC DTM (~100 m/px, Gwinner et al., 2005) topography shows a small bulge (Fig. 1d) which is much less pronounced than that observed in **the** Firsoff **crater** (Fig. 1b, c, d). The MU unit is present in patches, and mounds are particularly concentrated within this unit but also present also in the unit ELD although in lesser number. They are both grouped in the southeastern side and also towards the small central bulge of the crater (Fig. 1d). The mounds are sub-circular features with the same albedo of the ELDs, in stratigraphic continuity and most of the times presenting fine grained texture while showing an apical depression in only a few cases.

The "unnamed" crater, 20 km eastern of Firsoff, is the smallest of the three (~40 km), with a relatively smooth floor, scattered with several knobs and mounds that in places appear to be aligned along concentric fractures, contouring a subtle bulge visible from HRSC DTMs (100m) (Fig. 1d). Most of them seem to be well-preserved, although several appear slightly drop-shaped with a sharp crest due to wind erosion. An apical vent is not seen in either of these structures. **H**owever, from HiRISE observations, these structures have strong similarities in terms of texture and morphology to those mapped by Pondrelli et al., (2011) as well as to the indurated mound-like outcrops shown in **the southern** Vernal crater by Allen and Oehler (2008) and to the mounds within the hypothesized hydrothermal system in **the** Auki crater by Carrozzo et al. (2017).

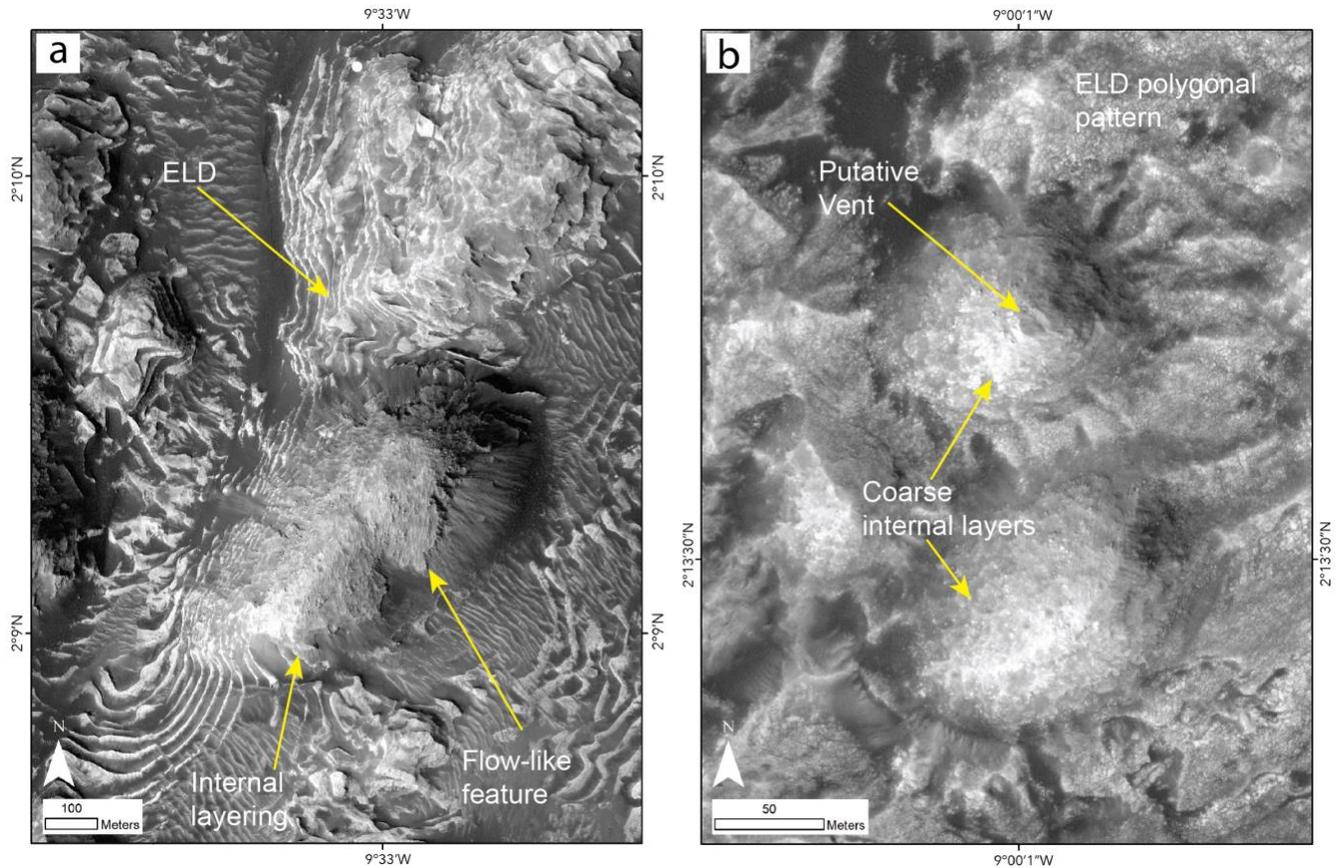

**Fig.2**: High-resolution close ups of putative mud volcanoes in the Firsoff Crater with referenced characters highlighted according to the interpretations in Pondrelli et al., (2011) and references therein. In a) Composite edifice with boulder rich flanks, with alternating fine-grained and boulder-supported material. Internal coarse layering appears in continuity with the surrounding ELDs. A flow-like boulder-rich formation covers the summit of the composite mound. In its NE part it is possible to spot an apical depression interpreted as the source of the flow-like feature. The albedo appears medium to low. In b) three single boulder-supported mounds presenting light-albedo. The northernmost edifice presents a barely visible coarse internal layering and an apical vent. It is clear the distinction in texture between the coarse-breached mounds and the surrounding ELDs that present a polygonal pattern. HiRISE images, 0.25 m/px.

## 2.3 Mounds within craters

Small mounds and their relationship with ELDs were first described in detail by Pondrelli (2011, 2015) within Firsoff crater, which presents an inner bulge mainly composed by ELDs. The mounds consist of simple cones (sub-circular or slightly elliptical, 50-300 m in diameter and 20-120 m height), and coalescent cones (500 m in diameter and several hundred meters high). In some cases, a main circular body and a sub-circular secondary appendix are also visible (Figs. 2a, b; and figures in Pondrelli et al., 2011). At HiRISE resolution, the mounds appear to have a fine-grained matrix-supported texture with meter-size light-albedo boulders, whose occurrence is higher at their base, Fig. 2a, b), and in some other cases a coarser boulder-supported texture. Larger mounds (~100 m in height) show boulder-rich layers alternated with fine-grained layers, sometimes slightly outward-dipping. In these latter cases, a layering **made of coarse breccia** is often in clear continuity with ELDs, suggesting a common genetic process related to fluid expulsion alternating periods of activity and quiescence (Pondrelli et al., 2011).

Depending on their position some of the mounds appear to be wind shaped, with a slightly sharp crest and an elongated shape in plan view. In general, the overall mounds' morphology and appearance also show striking similarities with the hypothesized spring mounds within the impact-induced hydrothermal system in Auki crater (Carrozzo et al., 2017) and with the terrestrial small mud volcanoes and gryphons in Dashgil (Mazzini and Etiope, 2017, Bonini, 20**12**, Bonini and Mazzarini 2010, Dimitrov, 2002).

*2.4 Evidences of mud volcanism and fluid expulsion associated to mounds*

Several interpretations have been provided in literature to explain the presence of such mounds within impact craters. One of the favored hypotheses relates mounds formation to aeolian differential erosion of lacustrine/aeolian sediments. The small mounds can be thus interpreted as a peculiar type of aeolian landforms know on Earth as whale-back yardangs. These landforms can be found on Earth in the Qaidam basin in China and their analogues in several places on Mars such as Arabia Terra and Gale (Xiao et al., 2017). Their shape can be scalloped and rounded on top, with a drop-shape in plan-view when isolated and more rounded shape when these features are smaller and more coalescent. Deposition in a lacustrine environment and differential erosion can cause the formation of such features as demonstrated by Heermance et al., (2013), however do not account for outward-dipping layers within the Martian mounds. Wet-dry cycle in depositional conditions is another factor that could play a role in the sediments induration and different attitudes of the layers, accounting both for the presence of intra-crater bulges and the mounds as a remnant of different depositional scenarios driven by climate changes on Mars (Kite et al., 2016). The cyclic transition between wet-dry conditions and subsequent erosion could cause indeed the formation of small ridges on the major intra-crater bulge (and the bulge itself) but nonetheless does not account for such a difference in shape and the more rounded mounds that are widespread in between the more elongated and drop-shaped yardangs.

Other models such as Zabrusky et al., 2013 and Bennett et al., 2016, relate the intra-crater bulge to aeolian activity and erosion but do not account for the presence of these mounds often aligned along ridges (Pondrelli et al., 2011, 2015). Moreover, such mounds are interpreted for spring resurgence in Pondrelli et al., (2011, 2015 and in Dimitrov, 2002; Kopf, 2002). The textural characters suggest the occurrence of poorly sorted material as in mud volcanoes sampling material from deeper levels intermixed and supported within a mud matrix (Pondrelli et al., 2011 and references therein).

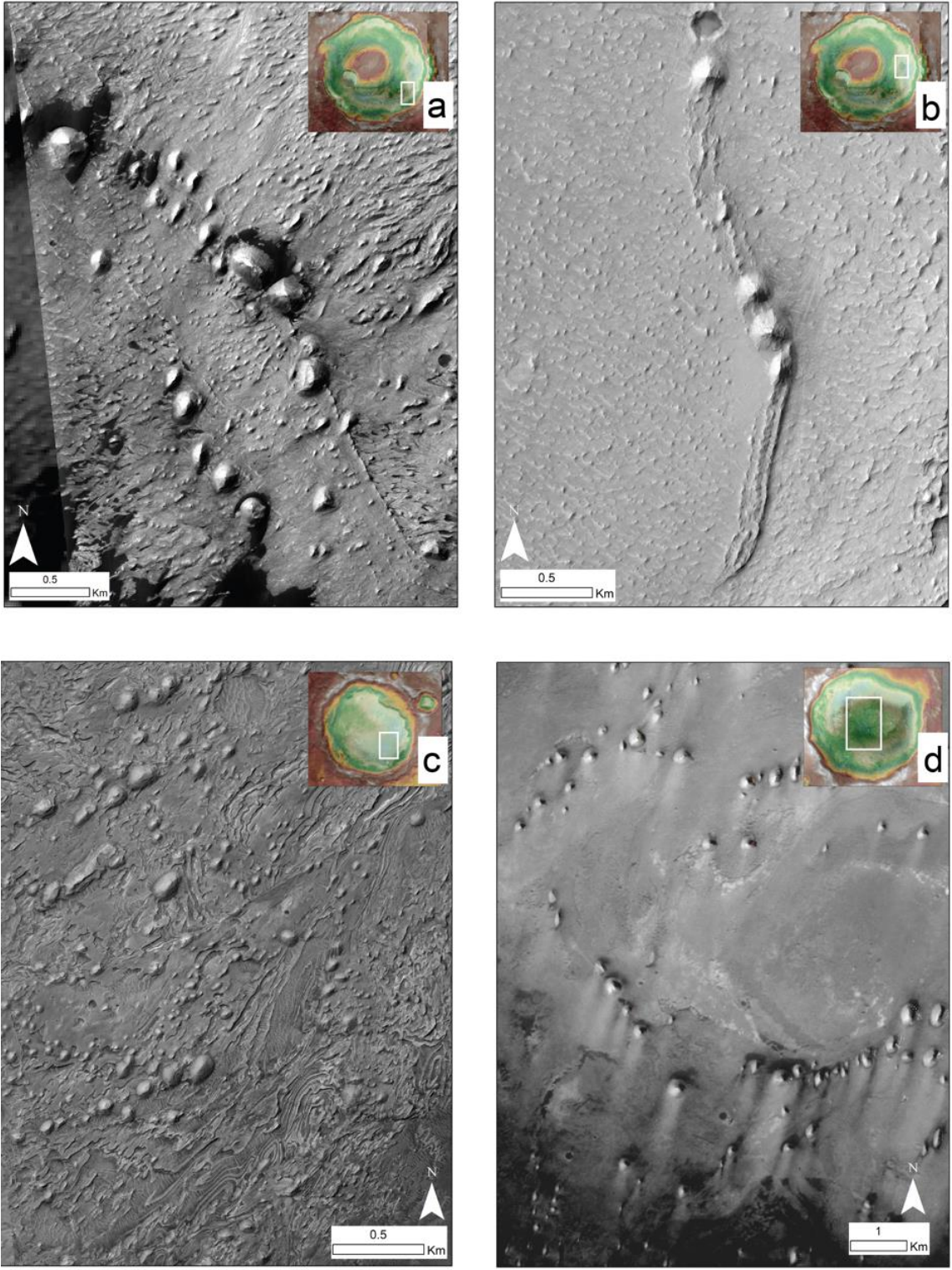

**Fig.3**: Aligned mounds in Firsoff with an a) radial trend and b) concentric trend. Similar alignments can be found in b) Kotido and c) eastern unnamed crater where the trend is concentric. The aligned mounds are placed along fractures and possible fissure ridges (e.g., Pondrelli et al., 2015).

Thus, although many other models have been proposed to explain the presence of intra-crater bulges and mounds, several pieces of evidence support the fluid upwelling-related origin hypothesis, linking the ELDs and the small mounds formation processes.

In the south-eastern part of the Firsoff crater, apical depressions on ~35% of the mounds were observed, especially on those mounds with boulder-supported coarse texture; such a large percentage of mounds with apical depression statistically rules out the impact origin of these depressions whose origin could, instead, be referred to an apical vent. Circular depressions on the flanks or at the base of the mounds could instead represent secondary vents. The lack of vents in the remaining mounds with similar overall shape maybe due to erosion or dust mantling, or to lower extrusive activity (Pondrelli et al., 2011).

In some cases, flow-like features are present on the mounds' flanks spreading from the apical depressions (Fig. 2a).

Whereas the majority of the mounds appear to be scattered, in several cases, they are observed along alignments interpreted as tens of meters-high fissure ridges caused by fluid extrusion and hardening (Pondrelli et al., 2015). In fact, mound populations and in particular the alignments (Fig. 3), show strong resemblance with the fluid expulsion inundated outcrops identified in the near Vernal Crater (Allen and Oehler, 2008).

Spectral signatures on ELDs and mounds showed the prevailing presence of polyhydrated sulphates and hydrated phases, (Pondrelli et al., 2015, Grotzinger et al., 2005) implying the presence of groundwater similar to those observed in the nearby Meridiani Planum (Flahaut et al., 2015).

We analyzed our case study according to this interpretation, which is the more comprehensive for mounds morphology, spatial arrangement and composition. In the following sections we present an automated mapping extraction workflow for the above described features in order to assess their possible fractal clustering and the geologic implications.

**3 Datasets used for mounds automatic detection**

In the Firsoff crater, mounds were first interpreted as mud volcanoes by Pondrelli et al., (2011; 2015). The morphometric features of these mounds (Pondrelli et al., 2011 and 2015) are used as a reference in order to map and analyse putative mud volcanoes within the three analysed craters. To do that, we used HiRISE (High Resolution Imaging Science Experiment, McEwen et al., 2007) data providing an image resolution of 0.25 m/pixel and stereo-derived DTMs with 1m resolution.

| Crater | Instrument | Stereo pairs product IDs | Pixel Scale (m/pixel) | Phase Angle | Convergence Angle | Vertical Precision (m) |
|---|---|---|---|---|---|---|
| Kotido West | CTX | F05_037663_1794_XI_00S009W<br>G10_021945_1800_XI_00S009W | 5.43<br>6.05 | 26.83°<br>48.38° | 19.8° | 3.0 |
| Kotido East/Firsoff South-East | CTX | B18_016776_1818_XN_01N009W<br>B19_016921_1818_XN_01N009W | 5.46<br>5.47 | 41.84°<br>56.96° | 18.5 | 3.4 |
| "eastern" crater | CTX | D18_034129_1822_XN_02N007W<br>F18_043017_1824_XN_02N007W | 5.39<br>6.07 | 43.02°<br>25.91° | 25.3° | 2.4 |
| Firsoff | HiRISE (used for reference in Firsoff) | PSP_003788_1820_RED<br>ESP_020679_1820_RED | 27.2<br>28.2 | 54.4°<br>69.2° | 15.1° | 0.2 |

**Table 1**: Details of stereo pairs used for DTM generation within the three craters and stereo matching parameters. The best CTX stereo-paired images and the DTM vertical precision was calculated with the PILOT stereo-matching tool (Bailen et al., 2015).

Since HiRISE observations are targeted only to very specific areas, we aimed to extend our mounds detection to larger areas covered by CTX camera (Context Camera, Malin et al., 2007). With a ground resolution of 6 m/pixel and a swath width of 30 km and a variable length of 50-300 km it, the mosaic of two of its images encompasses completely the craters, and the overlaps allow to generate stereo DTMs (18m resolution).

The PILOT stereo tool automatically selects several combinations of overlapping CTX images for stereo DTM reconstruction on the three craters (see table 1 for image details). Pre-processing was performed by means of USGS ISIS3 software suite. This was used to calibrate, de-stripe, and map project the images. DTMs were generated using ISIS3 (Integrated Software for Imagers and Spectrometers, Torson and Becker, 1997) and ASP (Ames Stereo Pipeline by Moratto et al., 2010; Beyer et al., 2014; Shean et al., 2016) following the procedures and using wrapper scripts by Mayer and Kite, (2016). Bundle adjusted CTX point clouds, obtained from stereo matching, were aligned to MOLA (Mars Orbiter Laser Altimeter onboard Mars Global Surveyor, Smith et al., 2001) PEDR (Precision Experiment Data Records) shots using iterative closest point algorithm and interpolated to a DTM with 18 m resolution. As an additional product, orthorectified CTX images at 6 m/pixel were also generated.

Among the generated DTMs by CTX overlaps proposed by PILOT, we selected those with less artifacts and no-data regions (thus providing the best stereo reconstruction) which resulted to be generated from the overlapping CTX pairs with convergence angle comprised between 15 and 25 degrees (see Table 1 for details).

The such obtained DTMs, 2 for Firsoff and 1 for Kotido, cover almost entirely the craters. However, in the case of the eastern crater the desired quality of the CTX DTMs was not sufficient in some areas, and therefore we integrated also HiRISE frames provided by the HiRISE team webpage and re-interpolated to create 18m point spacing DTMs (matching the CTX DTM resolution).

All the datasets were finally projected in sinusoidal centered on each crater and used for further analysis and display.

**4 Methods**

*4.1 Mounds automatic detection*

Even though it is still possible to identify most of the mounds at CTX image resolution a clear distinction between mounds and erosional features such as mesas and yardangs is often ambiguous and unreliable.

In this study, a supervised morphometric approach based on DTMs is adopted in order to distinguish between mounds from yardangs and ELDs erosional remnants. In order to achieve this, it was important to first calibrate the analysis on a well-studied site. We extracted the morphometric parameters of the small mounds mapped in **the** Firsoff crater that were interpreted as mud volcano candidates in Pondrelli et al., (2011, 2015). These were used as a base line for further mound extraction in less known areas (i.e., the Kotido crater, the unnamed crater and locations in the Firsoff crater not covered by HiRISE data) by using a HiRISE DTM resampled at CTX DTM resolution (18m).

The TPI (Topographic Position Index, Weiss, 2001; Jenness 2006) is the basis of our morphometric classification (see also Appendix A.1). TPI relies on the difference of elevation between a cell and the average elevation of its neighborhood **returning an index which is dimensionless**. TPI along with the cell slope value can be used to classify the cells into classes related to different specific morphologies (hills, narrow valleys, plains, etc., Jenness et al., 2006). For example, TPI-based techniques were used to automatically extract pockmarks in the Barents Sea (Gafeira et al., 2018), to map deep seafloor mounds in the Canary Islands basin (Sanchez-Guillamòn et al., 2018) and live biogenic reefs in Mingulay Reef Complex in Scotland (De Clippele et al., 2017). In all these cases the topography surrounding the analyzed features was relatively flat floored making them clearly distinguishable. In our case instead, we had to cope with a much more complex and variegated topography in the surroundings of the mounds, especially for the Firsoff crater. For this reason we used a multi-scale approach based on the combination of large and small DTM cell neighborhoods in order to combine small positive topographic expressions (i.e., mounds, set as 100 m treshold) within larger ones (i.e., the inner crater topography, set to 1000 m, Jenness et al., 2006). It is important to note that TPI is scale dependent. The presence of a small hill top within a narrow valley will be hidden if the chosen kernel size is larger than the valley itself and on the other hand, a hill top may not be visible if the window size is smaller than the hill itself. Hence the relation between the window size and dimensions of the analyzed morphological features must be taken into account, especially in such a topographically variegated terrain as in Firsoff.

| **Landform** | **TPI class** | **Morphology within craters** | **Curvature classes** |
|---|---|---|---|
| Narrow incisions | 0-2 | Mounds base/yardangs base | -0.0145 - -0.0032<br>-0.0032 - -0.002<br>-0.002 - -0.0009 |
| Gentle downslopes/ drainages | 3 | Equatorial layered Deposits/MV | -0.0009 - -0.0005 |

| Broad flat areas/plains | 4 | Equatorial Layered Deposits/MV | -0.0005 - -0.0002 |
|---|---|---|---|
| Broad open slopes | 5 | Mounds flanks | -0.0002 - 0.00009 |
| Upper slopes/mesas | 6 | Mesas/terraces | 0.00009 - 0.0005 |
| Hills in valleys/local ridges | 7 | Ridges | 0.0005 - 0.002 |
| Hilltops/Local ridge in plains/valleys | 8, 9 | Mounds/ Yardangs | 0.002 - 0.004 0.004 - 0.02 |

**Table 2**: Landform TPI classes according to Weiss (2001) and Jenness, (2006) and their correspondent feature within the three craters.

We were able to classify the multi-scale TPI output according to the classes identified in Weiss (2001) and are displayed in table 2.

The TPI values equal to or larger than 8 are those that identify small positive reliefs such as the mounds and yardangs crests.

However, to better constrain the mound morphologies and for the automatic mapping of putative mud-spring/mud-volcanoes, the TPI classification was used along with the *profile curvature*, that has been calculated on the same DTM using the *r.param.scale* GRASS module (Hofierka et al., 2009). The *profile curvature* is the curvature calculated along the maximum slope directions and is very sensitive to slope variations (Wood, 2009). The obtained values allowed assigning a specific range of profile curvatures, being convex, concave or flat (Fig. 4b) to the different TPI geomorphological classes. Moreover, we used the zero-profile curvature (corresponding to the point in the mounds slopes where curvature changes from concave at the base to convex towards the top) and its intersection with TPI< 8 to automatically contour the mounds in class 8-9 and gather the most realistic shape in plan-view. This method avoids any interference with possible topographic irregularities of the surrounding terrains, such as open slopes, plains with a certain degree of roughness or narrower valleys. (Fig 4c).

The mounds on the south-eastern sector of the Firsoff crater, as studied by Pondrelli et al. (2011), present positive profile curvatures between 0.002 and 0.004 and fall in the TPI category 8 and 9, whereas the sharp crests of yardangs fall uniquely within category 9 with curvatures between 0.004 and 0.02. In order to remove high frequency noise/artifacts, objects smaller than 50 m were filtered out based on the 4-pixel thumb rule also used in crater counting (i.e., a rounded object needs to be at least 4 pixel in diameter to be recognized). Yardangs, false positives, and larger artifacts in the DTM were instead filtered using aspect ratios in plan-view, which were obtained by extracting minimum and maximum axes for every contoured feature (see also Appendix A.1.2). All objects with an aspect ratio less than 0.5 (i.e., highly elongated features such as yardangs, crater rims, and ridges) were discarded. This minimum threshold was chosen in accordance to the mound aspect ratio calculated on the mounds mapped in Pondrelli et al., (2011) that display an aspect ratio equal or greater than 0.5. By applying these filters to mounds mapped in the same area used as Pondrelli et al. (2011), an automated data set was generated (see also Appendix A.1). In addition, since the mapped mound area is defined within the ELD and MU geologic units (Pondrelli et al., 2015), and some knobs or positive reliefs features resulted from the emergence of eroded strata banks we considered only the mounds within broad areas with slope gradient less than 15°. The automated data set and the original one by Pondrelli et al., 2011 show a strong correlation with the majority of the original mounds correctly mapped (see also Appendix A.2) (Fig. 4c, d). This method has been calibrated on a known Firsoff area using HiRISE DTMs resampled at the CTX DTM resolution (18m post-spacing). After testing, the above method was applied **to** the 2 CTX DTMs covering the Kotido and the unnamed craters respectively. We finally used 6m/pixel CTX orthoimages, and HiRISE single images (where available) for a visual checking of the results in order to evaluate the presence of still ambiguous objects.

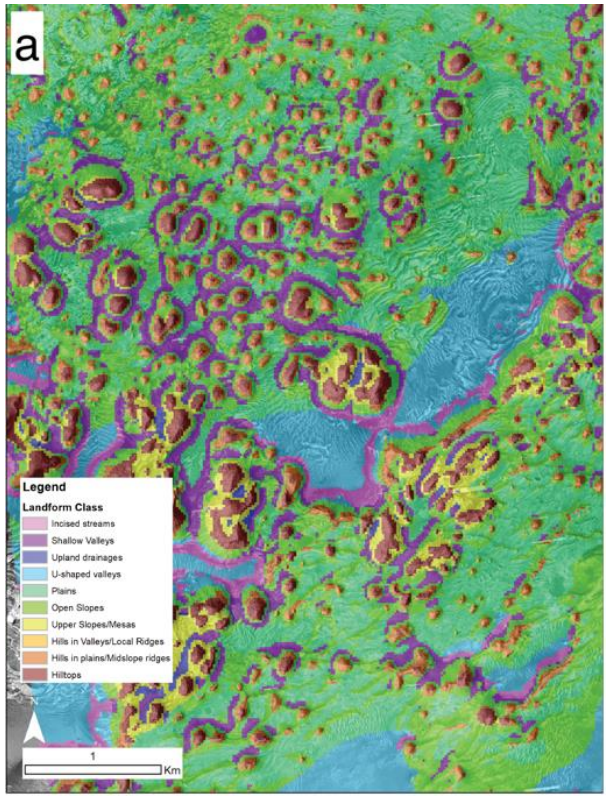 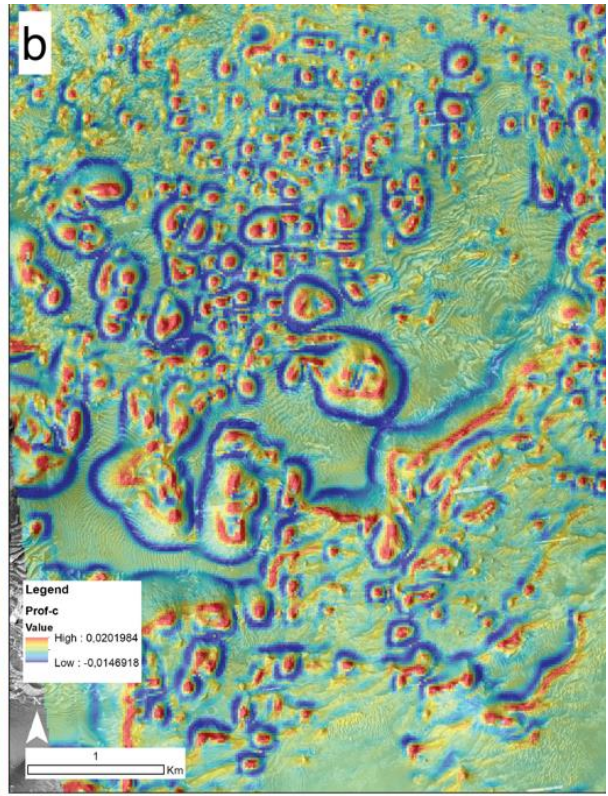 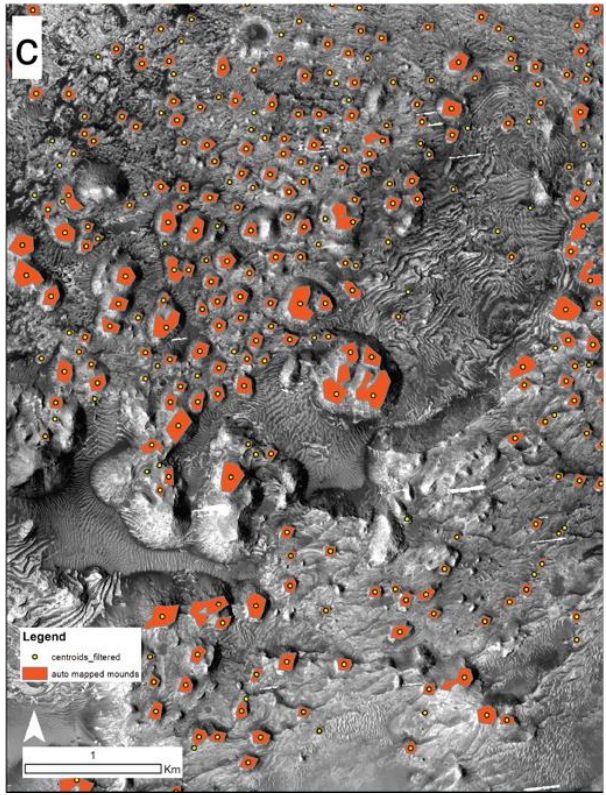 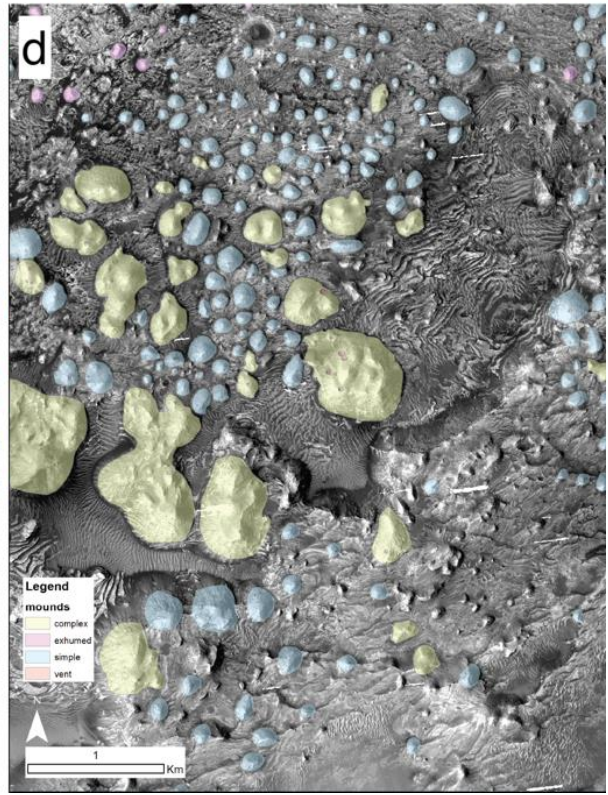

**Fig. 4**: a) TPI map with landform categories according to Weiss (2001) and Jenness (2006), b) Profile curvature extracted from DTM resampled at CTX DTM resolution of 18 m, c) automated contours of mounds extracted along zero-profile curvature and their centroids used for fractal clustering analysis d) mounds mapped by Pondrelli et al (2011) in the area marked in fig. 5a'.

The centroids of the contoured mounds, corresponding to the position of the putative mud volcano centers, were then extrapolated (Fig. 5) and studied in terms of self-similar clustering of their spatial distribution.

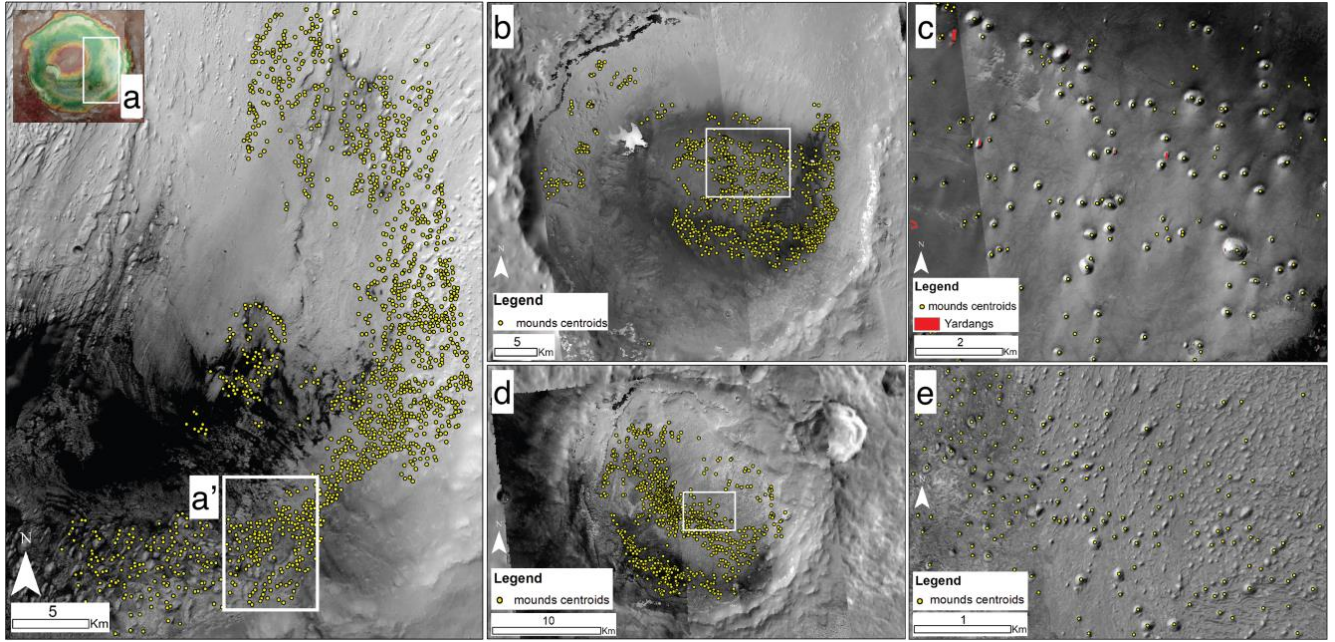

**Fig. 5**: Automatic mapping of mounds in a) South Eastern Firsoff with a') being the subset mapped in Pondrelli et al., (2011) and shown in fig. 4. Panel b) shows the whole dataset extracted in the unnamed crater with a detail of the mapping in c). In d) Kotido crater dataset with e) detail of the automatic mapping.

*4.2 Analysis of mounds' spatial distribution*

    The spatial distribution of monogenic vents in volcanic areas on Earth (Mazzarini and Isola, 2010) and Mars (Pozzobon et al., 2015) are linked to fracture systems that allow an efficient hydraulic connection between the surface and crustal/subcrustal fluid reservoirs. The percolation theory describes the geometric and physical properties of a percolating network (Stauffer and Aharony, 1992; Orbach, 1986; Song et al., 2005) and can be applied to fracture networks that serve as a pathway for fluids to move within the crust.

The strength of the self-similar clustering approach is that it applies to processes where fluids, whether magma or mud, are withdrawn from a reservoir at some depth (the source) and emitted to the surface (the sink) via a percolating network of fractures (i.e., Bonini and Mazzarini, 2010; Mazzarini and Isola, 2010). Indeed, among different features on Earth and Mars sharing similar morphologies, such as volcanic vent, mud volcano, pingos and tumuli, a clear self-similar clustering has been observed only for those clearly linked to a source to sink connection (i.e. volcanic vent and mud volcanoes; De Toffoli et al., 2018).

The first step in the analysis of spatial distribution of mounds/mud volcanic features in the Firsoff and nearby craters was the computation of the nearest neighbor distance (NN or point separation) for each data set. The clustering of data has been analyzed by computing the coefficient of variation (CV) and R-c test on the point separation values. The CV is the ratio between the standard deviation and the mean of the sampled population (Gillespie et al., 1999). A value of CV > 1 indicates point clustering, CV = 1 indicates a random or Poisson distribution, and CV < 1 indicates anti-clustering (a homogeneous distribution). CV investigates how close are the points are to each other, and gives information on short range clustering but not on the pattern of point distribution. R-c statistics (Clark and Evans, 1954) compare actual NN distance distribution with that expected for a Poisson distribution of N points. R < 1 indicates clustering. To identify statistically significant departures from randomness at the 0.95 and 0.99 confidence levels, |c| must exceed the critical values of 1.96 and 2.58, respectively (Clark and Evans, 1954). The reference density is obtained from the ratio between the actual point number and the area of the convex hull containing them (e.g., Baloga et al., 2007; Beggan and Hamilton, 2010).

The spatial distribution (self-similar clustering) of mounds has been investigated by applying the two-point correlation function method. For a population of N points (e.g. mounds within the crater), the correlation integral is defined as the correlation sum (C(l)) that accounts for all the points at a distance of less than a given length l (Bonnet et al., 2001; Mazzarini and Isola, 2010). The term is computed as

$$C(l) = 2N(l)/(N(N-1)) \qquad (1)$$

where N(l) is the number of pairs of points whose distance is less than l. The fractal distribution is defined by

$$C(l) \sim bl^D \qquad (2)$$

with *b* being the normalization constant and *D* the fractal exponent. The slope of the curve in a log(C(l)) versus log(l) diagram yields the D value. The computed D value (fractal exponent of clustering) holds for a defined range of distances (size range) where the equation is valid. For each analysis, the size range of samples is in turn defined by a plateau in $\Delta\log(C(l))/\Delta\log(l)$ (i.e., the local slope) versus $\log(l)$ diagram. The wider the range the better the computation of the power-law distribution (Walsh and Watterson, 1993). The derivation of the cutoffs bounding the size range is a crucial point and is generally not trivial, especially when the local slope does not show a regular and wide plateau (see also Appendix B). The choice of the zones where the plateau is well-defined and the determination of the lower and upper cutoffs (*Lco* and *Uco*, respectively) are done by selecting the wider length range for which the correlation between $\log(l)$ and local slope is greatest (Mazzarini, 2004). A size range of at least one order of magnitude and at least 150 samples is required to extract robust parameter estimates (Bonnet et al., 2001; André-Mayer and Sausse, 2007; Clauset et al., 2009). By analyzing the volcanic vent clustering (Mazzarini and Isola, 2010), it has been shown that the random removal of 20% of the analyzed samples from large datasets (i.e., >200 vents) does not affect the estimation of fractal dimension (less than 0.01% of variation) and the error introduced into the estimation of the cut-offs is less than 1%–2% (Mazzarini and Isola, 2010). Mazzarini et al., (2013) in order to test the effect of uncertainties in point-like feature locations added random errors to the sampled points (in the 0–100 m, 0–300 m and 0–500 m ranges, i.e., errors as high as 5 to 25 times that of the coarsest image resolution used to locate the points). The 0–100 m errors randomly added to the point (vent) locations generated fractal exponent and cut-off values identical to those computed for the original dataset. In the case of 0–500 m random errors, the resulting fractal exponent was 3% higher than that computed for the original dataset, and the cut-offs were very similar to those computed for the original dataset (Mazzarini et al., 2013). It has been shown that the upper cut off value (*Uco*) obtained analyzing several volcanic fields on Earth linearly scales with the depth of the fluid source (e.g., Mazzarini and Isola 2010). This relationship has been observed for volcanic vents located in different geotectonic settings (Table 3) such as in the East African Rift (Mazzarini, 2007; Mazzarini and Isola, 2010), southern Patagonia (Mazzarini and D'Orazio, 2003; Mazzarini et al., 2008), TransMexican Volcanic Belt in Mexico (Mazzarini et al., 2010) and for mud volcanoes in the Greater Caucasus in Azerbaijan (Bonini and Mazzarini, 2010). The best linear fit between the computed Uco values and the depth of the fluid reservoir (T) is Uco = 0.98T − 0.6 with $R^2$=0.95; errors are 20% for T derived from independent geophysical data sets and 10% for Uco estimates (Fig. 6, and Table 3).

| Data set | n | D | Lco (km) | Uco (km) | $R^2$ | T (km) | Geodynamic setting | References |
|---|---|---|---|---|---|---|---|---|
| South Afar | 134 | 1.19±0.04 | 1.7±0.8 | 22.8±1.5 | 0.99 | 23 | Extensional | 1 |
| Central Afar | 203 | 1.52±0.02 | 1.2±0.2 | 11.5±0.8 | 0.99 | 11 | Extensional | 1 |
| North Afar | 349 | 1.34±0.02 | 1.4±0.4 | 14.2±2.3 | 0.99 | 15 | Extensional | 1 |
| MER | 391 | 1.17±0.02 | 2.8±0.3 | 10.1±1.4 | 0.99 | 10 | Extensional | 1,2 |
| Virunga | 287 | 1.50±0.01 | 1.0±0.1 | 21.6±3.1 | 0.99 | 20-30 | Extensional | 2 |
| Michoacan–Guanajuato | 923 | 1.67±0.02 | 1.3±0.2 | 38.1±3.2 | 0.99 | 40-42 | Transtensional | 3 |
| Sierra de Chichinautzin | 181 | 1.56±0.02 | 1.5±0.4 | 32.0±3.7 | 0.99 | 30-50 (>40) | Transtensional | 3 |
| Pali Aike | 467 | 1.45±0.02 | 0.8±0.2 | 34.0±2.0 | 0.99 | 32 | Back-arc | 4 |
| Payen | 675 | 1.35±0.01 | 0.8±0.4 | 47.1±3.6 | 0.99 | 45-50 | Compressional | 5 |
| Arzebaijan | 480 | 0.52±0.01 | 0.3±0.1 | 5.1±0.5 | 0.99 | 5-7 | Compressional | 6 |

**Table 3**: For each location dataset here are displayed the number of vents (n), the fractal exponent *D*, lower and upper cutoffs (Lco, Uco) with the relative $R^2$, and the depth of fluid source T obtained by independent geophysical datasets. The linear relationship between Uco and depth of the fluid source (T) scaling has been tested in different geodynamic settings in the penultimate column to the right. The references in the last column to the right are: 1) Mazzarini, 2007; 2) Mazzarini and Isola, 2010; 3) Mazzarini et al., 2010; 4) Mazzarini and D'Orazio, 2003; 5) Mazzarini et al, 2008; Mazzarini and Bonini, 2010.

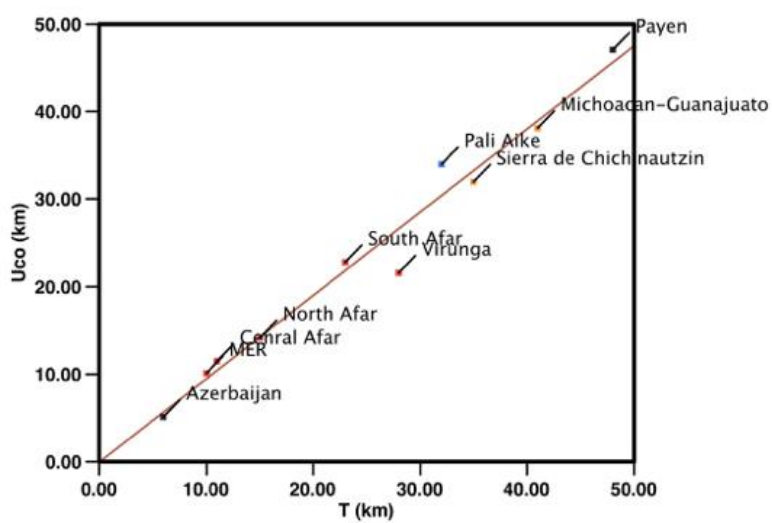

**Fig. 6:** Plot showing the relationship between the Uco calculated in different geodynamic environments on Earth and the depth (T) of the fluid source (magma,

mud) derived by independent geophysical datasets. It appears clear that the relation is almost linear presenting well-matching correspondence with the results derived by the fractal clustering and geophysical analyses.

## 5 Results

In the case of the mounds that were automatically extracted, it appears that those in the craters that show asymmetric distribution are more frequent in the upstream side (SE-NW, referred to the regional hydraulic gradient and slope) of the Firsoff crater. In Kotido they appear mostly concentrated in the middle of the bulge area, whereas in the unnamed crater mounds are more homogeneously scattered (Fig. 5). The results from the fractal clustering analysis on the mounds distributions are as follows:

Firsoff crater (N=1578) has average NN distance 0.51 km, CV = 1.44 and R-c statistics of 0.34 and -26.2; eastern unnamed crater (N=1036) CV = 1.61 and R-c statistics of 0.57 and -26.42; finally Kotido crater (833) has CV 0.65, R-c statistics of 0.79 and -5.6. Firsoff craters display both short and long-range clustering whereas the Kotido crater, where a well-defined central bulge is not present, shows clustering only at large scale. The unnamed crater shows fractal clustering with a well-defined plateau. In table 4 the obtained fractal clustering results on the datasets from the three craters are summarized. The analysis of mounds clustering in the Firsoff crater suggests a source depth ($Uco$ value) of 2.6±0.3 km from the crater's floor (both in the subset and the broader area, Fig. 5a, a', Fig. 7 a, b). We chose to maintain the analysis in south-eastern Firsoff mounds as they were more preserved from erosion and with clear evidence of alignments along ridges. Similar results (Fig.7c, d) were also obtained also from the fractal analysis of the mounds in the unnamed crater ($Uco$ = 3.2 ±0.4 km) and for mounds in the Kotido crater ($Uco$ = 2.7 ±0.3 km) (Fig. 6c, Fig. 5b, c, d , e, and Appendix B). The Lco and Uco are calculated according to the method described in Mazzarini (2004): for both of them we selected the wider length range for which the correlation between log(l) and the local slope is greatest. The actual fluid depth derived by the analysis of mound self-similar clustering (Uco) is referred to the elevation of mapped mounds. To evaluate the actual depth of the pressurized fluid table (H) feeding the mounds we must add to the Uco the difference in elevation between the mounds and the surrounding plains (Δh) to obtain H=Uco + Δh (see table 5 for further details).

| Dataset | n | D | $R^2$ | Lco | Uco |
|---|---|---|---|---|---|
| Firsoff (Pondrelli et al., 2011) | 259 | 1.3932 | 0.9996 | 0.9±0.2 | 2.6± 0.3 |
| Kotido Crater | 833 | 1.7593 | 0.9999 | 1.3±0.2 | 2.7±0.3 |
| SE Firsoff | 1578 | 1.7845 | 0.9999 | 0.8±0.2 | 2.6±0.3 |
| Unnamed | 1037 | 1.6885 | 0.9999 | 0.4±0.1 | 3.2±0.4 |

**Table 4**: Values derived from the fractal analysis. The Uco, which correspond to the depth of the fluid reservoir calculated from craters' floors varies between 2.6 and 3.2 km. The difference in the fractal exponent D value between south-eastern Firsoff and its subset (Pondrelli et al., 2011) is probably due to the large difference of the mounds populations within each analyzed crater. However, the South Eastern Firsoff population is highly compatible with the D values from the other two analyzed craters.

The pristine depths of the three craters are needed in order to assess if the fluid table is nested within the layered unit or below the crater, and therefore likely related to a pre-existing setting. Forsberg-Taylor et al. (2004) provides estimates of crater degradation in terms of diameter increase due to mass wasting, faulting and collapse of the inner walls (10% of diameter increase) and basin infilling (up to 2/3) caused by airfall, and aeolian and/or fluid erosion. Using these results, it is possible to estimate the most likely pristine crater diameter.

The pristine depth of the craters was calculated using the equations from Robbins and Hynek, (2013) that derived the morphometric relationship between complex craters diameter versus depth over different terrains comprised between 40°S and 40°N on Mars. By applying this approach, we derived the depth of excavation at the time of the impact (Table 5) and provide constrains for the possible thickness of the inner deposits. Table 5 highlights the resulting Uco from all the case studies and shows that they are located beneath the pristine depth of the craters placing the fluid source in each crater below its floor.

## 6 Discussion

Fluids and water-related activity on an ancient Martian surface has been described by several authors (Bennett and Bell, 2016, Zabrusky et al., 2012; Andrews-Hanna et al., 2010; Andrews-Hanna et al., 2011b; Michalski et al., 2013b; Grotzinger et al., 2008; Flahaut et al., 2015) based on the occurrence of alteration minerals from water-rock interactions and on stratigraphic evidences

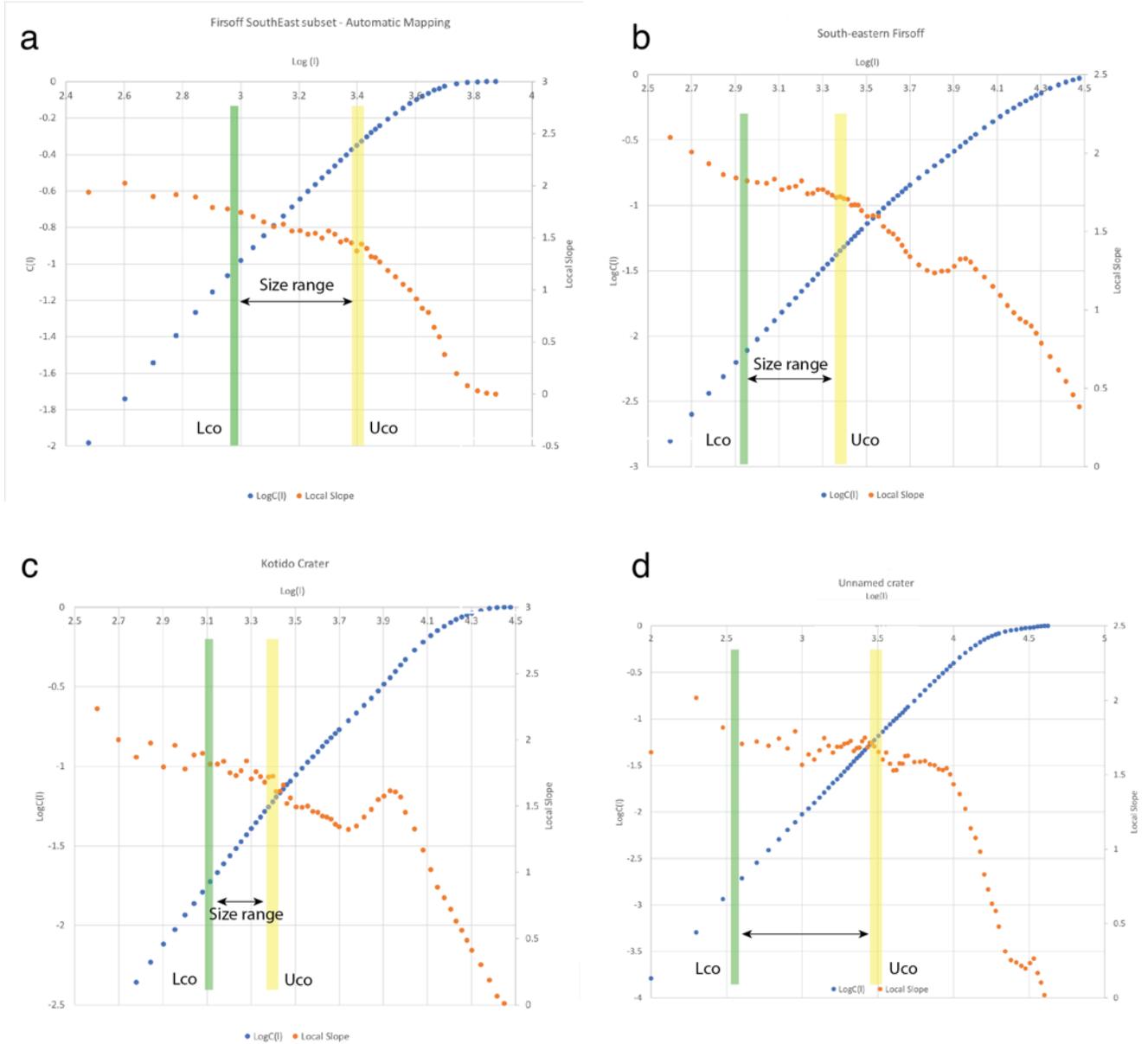

**Fig. 7**: Self similar clustering analysis for each dataset marked in fig. 5. The log plot of *C(l)* vs.*l* refers to equation (2) in the text. In the panel (a) it is shown the self similar clustering of mounds in the south-eastern Firsoff subset (i.e., fig 5a' and described more in detail Pondrelli et al. 2011) and in panel (b) the self-similar clustering in the broader

area in eastern Firsoff (shown in fig. 5a). In panel (c) it is shown the self-similar clustering of mounds in the Kotido crater and in panel (d) for the eastern unnamed crater. The slope of the line with blue dot is the fractal coefficient *D* in equation (2; see main text). The orange dots represent the local slope of the dataset (that is *Δlog(C((l))/Δlog(l)*, where *C(l)* is from equation (1) and *(l)* is the distance between points; see the main text) whose plateau represents the fractal behavior of the analyzed system in the size range defined by the colored bars in the log(l) vs. local slope plot. The thickness of the green and yellow bars represents the uncertainty in determining the Lco (lower cutoff) and *Uco* (upper cutoff) of the size-range distribution where *D* values show a plateau. The rightmost portion of the chard after the Uco yellow bar shows a drop of the *D* value indicating that the relationship between *C(l)* and *l* in log-log space is not linear, although with a high $R^2$.

| Crater | D | PD (km) | $d_R=0.28PD^{0.59}$ | Δh (average) | Uco | H=Uco+Δh |
|---|---|---|---|---|---|---|
| Firsoff | 80 | 72 | 3.5 | 1.20 | 2.60 | 3.80 |
| Unnamed | 40 | 36 | 2.3 | 1.00 | 3.20 | 4.20 |
| Kotido | 40 | 36 | 2.3 | 0.60 | 2.70 | 3.30 |

**Table 5**: Table representing parameters related to pristine depth and width of the craters along with the calculated fluid source depth with the fractal clustering analysis. The columns represent

D: actual diameter, PD: pristine diameter and d(r): pristine depth calculated according to Robbins and Hynek (2012), Δh: difference in elevation between the areas within craters presenting mounds and the outer plains, Uco: upper cutoff of fractal clustering corresponding to the fluid source depth. H: the depth of the calculated fluid table with respect to the surface calculated as the average plains elevation around the craters.

of large-scale spring deposits forming bulges within impact craters (i.e., Pondrelli et al., 2015; Franchi et al., 2014; Rossi et al., 2008). Indeed several numerical simulations were carried out (Andrews-Hanna et al., 2010, Andrews-Hanna and Lewis, 2011) to test the hypothesis of the deposition of ELDs as a consequence of fluid table oscillations through time. This repeated process would have filled the craters with sediments and evaporitic material creating layered sequences interbedded with fine aeolian material deposited during periods of quiescence. However, this model accounts solely for the presence of inverted craters in southern Arabia and Meridiani Planum and simple flat-floored, "layer cake" crater infilling, but does not clearly explain the presence of the large crater bulges often higher than the crater rims (such as Firsoff) and formed by symmetric outward-dipping layers.

The pedestal craters, inverted craters and intra-crater mounds were used to model a pre-erosional depositional surface which implies an average thickness of sediments of 6.2-11.6 m (Zabrusky et al., 2012). If the bulged craters would be explained by focused wind erosion, this thickness value, averaged on the whole Arabia Terra and Meridiani Planum, would imply an unreasonably uneven sediment removal by wind erosion on the region, being more than 1 km within some craters and very weak within other craters of the same age and in the surrounding plains. In fact, although aeolian depositional and erosional processes were numerically modelled to explain the presence of large intra-crater bulges and such a difference in height (as in the case of Mt. Sharp in Gale crater, Kite et al., 2014), it has been shown that, at least within this particular area of Mars, the aeolian or lacustrine deposition hypothesis is unlikely (Lewis and Aharonson 2014).

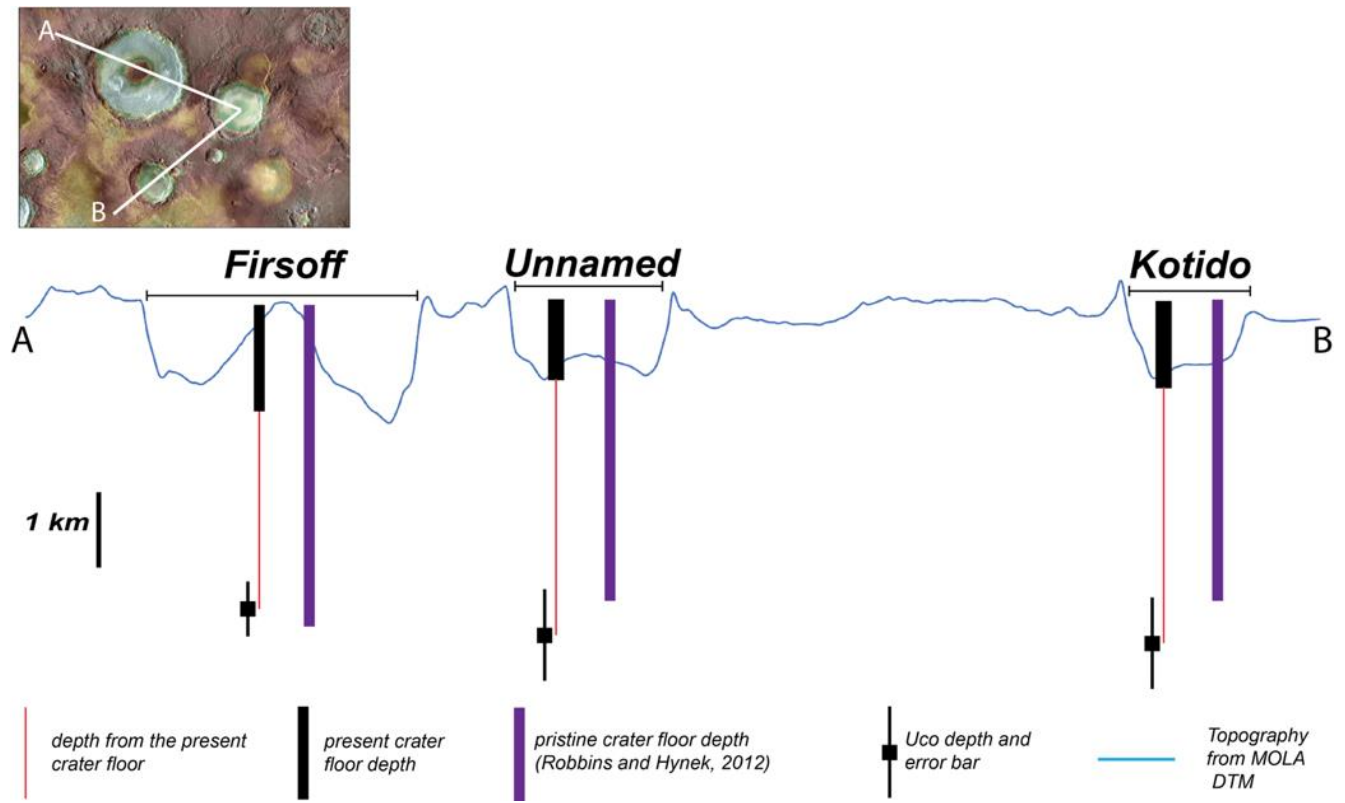

**Fig. 8**: Interpretative sketch showing the depth of fluid source feeding the mounds within the three craters. The inset shows the trace of the profile over the MOLA MEGDR (Mission Experiment Gridded Data Records) topography (Smith et al., 2001).

Moreover, the presence of small mounds scattered and sometimes aligned along ridges within several craters remains unexplained by both the water table oscillation and the focused wind erosion. Thus, another mechanism of formation, different from the simple interplay of lacustrine and aeolian ones must be invoked. In fact, if we consider such a strong wind erosion within crater basins, mounds would have been strongly shaped and be indistinguishable from yardangs. Instead, among yardang fields (e.g., south-eastern Firsoff) our mapping evidences a variety of fresh rounded morphologies pointing towards pristine uneroded forms. The presence of the mounds' alignment along fissure ridges (Pondrelli et al., 2011, Pondrelli et al., 2015; Franchi et al., 2014 and this work) and their morphological similarity with terrestrial mud volcanoes (Mazzini and Etiope, 2017), suggests that their origin is more likely related to fluid upwelling along fractures (Franchi et al., 2014; Rossi et al., 2008, Allen and Oehler, 2008). This observation is corroborated by the morphological similarities with putative spring mounds within the Auki crater that present striking similarities with the studied mounds in terms of size, morphology and textural characters (Carrozzo et al., 2017).

In addition, the mound alignments seem not to be randomly oriented, but instead follow a concentric pattern within the craters, particularly well-visible in the eastern unnamed crater (fig. 3d). This pattern of mounds along ridges is in agreement with the fractures arrangement of the Meteor Crater (Arizona) (Krumar and Kring, 2005). Here it is shown how within and around an impact crater an interplay of concentric and radial fracture patterns can be observed. In our case-study craters, the largest fractures could likely have been exploited during the fluid upwelling process leaving as a remnant a fissure ridge with elongated mounds. In Firsoff, the alignments are less common, although we could see a clear alignment radial to the crater (fig. 3a) and another one concentric (fig. 3b). In Kotido the alignments trend NE-SW slightly following a concentric pattern (fig. 3c).

A series of evidences suggest that mounds are intra-formational with ELDs (Pondrelli et al., 2015) and likely a manifestation of subsurface fluid upwelling such as the mounds. The occurrence of aligned mounds is coherent with the assumption of a fracture network beneath the mound field allowing the uprising of fluids from depth. The self-similar approach we perform focuses on the spatial clustering of mounds and further specific investigation will be carried on the structural pattern of "feeding" fractures.

The self-similar clustering analysis of the mounds distribution have shown that the fluid source is 3.2 to 4.2 km deep. The comparison between the pristine depth of each crater with the corresponding depth of the water table (H) obtained with the fractal clustering shows that the source underneath these craters is located from 300 m to 1.9 km from the pristine crater's depth (Fig. 8 and table 5). The minimum conditions needed to obtain fluid pressurization and expulsion can be inferred from the current average difference in elevation of 0.6-1.2 km of areas where mounds are present in Firsoff, Kotido and the unnamed eastern crater with respect to the surrounding plains. Considering the loss of overburden due to impact crater excavation a vertical load drop ρgΔh (where ρ is the density of crustal rocks and Δh is the pristine depth of the crater) can be foreseen. Fluid overpressure and hydrofracturing likely occurred when $\Delta Pf > T$, with T the tensile strength of the rocks. The 6.7-13.4 MPa pressure drop is computed assuming a range of Δh between 0.6 and 1.2 km and an average density of basaltic crustal rocks of ~3 g/cm$^3$, with tensile strength varying in the range 0.2-17 MPa depending on the integrity of the rock volume (Schultz, 1993; Jaeger et al., 1979). The basaltic rocks in the floor of the craters are likely highly fractured from the impact, and hence with tensile strength less than 5 MPa. Nevertheless, the computed ΔPf is likely to have provided sufficient load to overcome the tensile strength favouring the opening of fractures also within the subsequently deposited ELDs. Hence, we can infer that such overpressure conditions could have been even larger in the past, with higher Δh, favouring the hydrofracturing at the initial stages of ELD deposition when the crater infilling process was still active. The hypothesis of expulsion of pressurized fluids and related sulphate material deposition is also supported by the bulge elevation, the latter being slightly higher than the rim, as in the case of Firsoff and other nearby large craters such as Crommelin.

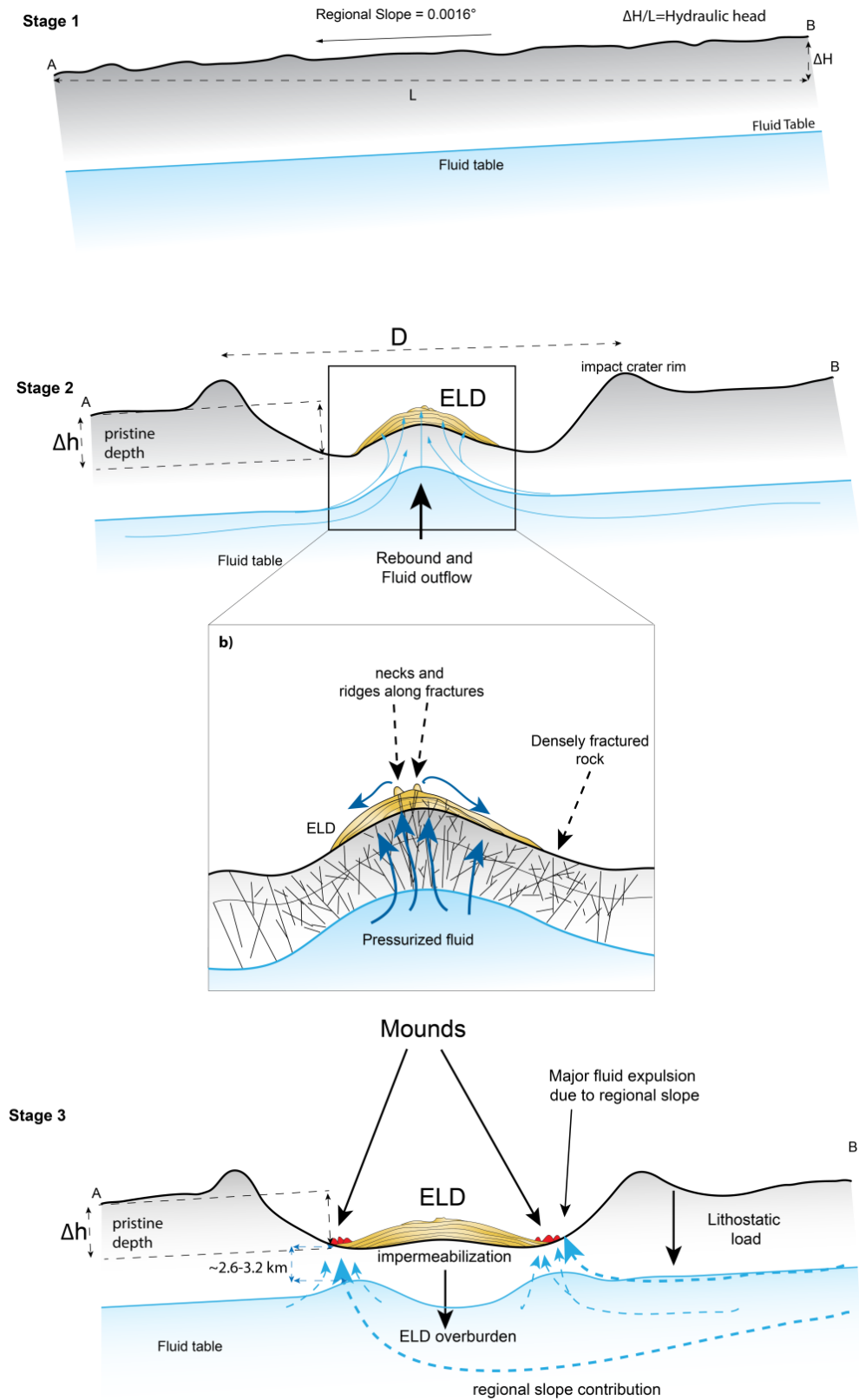

**Fig. 9**: Proposed evolutionary stages of formation of ELDs and mounds. See main text for further details.

We think that is very unlikely that the eolian processes where able to create the localized crustal unloading needed for triggered the fluid upwelling which is instead guaranteed by the instantaneous process of impact excavation. In addition, the compositional and genetic link between mounds, fissure ridges and layered deposits of the bulge (Pondrelli et al., 2015) rules out any aeolian deposition hypothesis and substantial effect in genereting the ELDs of the central bulge. Hence the aeolian surface modification would have taken instead place only at a later stage by sculpting already emplaced units giving rise to lower scale morphologies such as yardangs or the deposition of very fine dark layers interbedded within the ELDs (Pondrelli et al., 2015).

From all these observations we propose the following main evolution stages that led to mound formation within impact craters (Fig. 9):

1. Before the impacts a fluid table is present between ~3.2 and ~4.3 km underneath a still undisturbed surface (Fig. 9, stage 1)

2. Impact cratering phenomenon creates a sudden unloading and a consequential fluid overpressure. The pervasive fracture network, produced by the impact below and around the crater, is exploited by the over-pressurized fluids. This event, can additionally contribute in opening pre-existing sealed fractures further facilitating the fluid upwelling. Over-saturated deep fluids that are now able to reach the surface undergo sudden evaporation and precipitation due to the low atmospheric pressure and, as a consequence, start depositing ELDs. Because this upwelling process is directly linked with the amount of overpressure, we can hypothesize a decreasing of the entity of the upwelling through time. Fluids piped within the fracture system gave origin to mounds in the source locations (Fig. 9, stage 2). In the early stages, given the considerable thickness of the ELDs, the formed mounds can be hypothesized of large sizes and involving large amounts of fluids, with diameters up to tens of km resembling those described by Allen and Oehler (2009) in the nearby Vernal crater. As pointed out in Pondrelli et al., (2015), the mounds associated with this early-stage event are indeed in stratigraphic continuity with the ELD layers and show the same composition.

3. The progressive accumulation of the ELDs within the crater increased the overburden with a consequent decrease of permeability and sealing of the fractures in the ELDs main body (the inner bulge, Fig. 9, stage 3). The effect is that the pressurized fluids would now tend to flow out around the perimeter of the central ELD mound where the overburden is thinner (Fig. 1b, c) giving origin to the annular mounds distribution. On Earth, a similar mechanism involving sediments compaction and permeability inhibition influencing the localization of extrusion cones by creation of focused fracture pathways is also proposed by Allen and Oehler (2013). The progressive decreasing of overpressure through time gave origin to smaller mounds with coarser brecciated texture as a latter event. Moreover, the annular mounds distribution is not observed where a central ELD mound shows a subtle topographic expression or is not present at all as in the unnamed crater and Kotido, respectively.

In conclusion, the different distribution of the mounds and the diverse topographic expressions of the inner bulges can be I) related to different stages of the same process of fluid upwelling due to impact excavation and/or II) controlled by the size of the impact itself that can be also responsible of the duration of the process of over-pressurized fluids outpouring. At the moment, we can not unambiguously affirm that the mechanism of sediment extrusion is favored also by gaseous phases, however the mound morphologies are indeed well compatible with terrestrial mud volcanoes due to the presence of some flow-like features (fig. 2a) and several apical vents (Fig. 2a,b); this might even suggest a gaseous contribute..

**7 Conclusions**

The occurrence of a water table of regional extent in Arabia Terra as well as a fracture network required to mobilize pressurized fluids upward was suggested by several authors to explain the layered deposits within and outside the craters (Zabrusky et al., 2012; Michalski et al., 2013b; Andrews Hanna et al., 2010, 2011; Allen and Oehler, 2009, Allen et al., 2013; Rossi et al., 2008; Franchi et al., 2014).

Our results based on the mounds fractal clustering are consistent with the presence of a pressurized water table at ~2.6-3.2 km of depth and interacting within large impact craters such as the Firsoff, Kotido and an unnamed crater 20 km towards the east. The aquifer pressurization may have likely initiated by overburden removal due impact excavation, that produced pervasive fracture network. The pressurized aquifer and the exploitation of the fracture network by fluids is likely to have played multiple roles: I) generating depositional spring-related features such as large spring mounds and ELDs, II) producing hydrofracturing both within basaltic bedrock and ELDs themselves, and III) mud volcanism or small spring resurgences whose evidence are rounded necks as inferred from both morphological and compositional continuity with ELDs. As a consequence, similar subsurface fluid flow processes could be expected in other craters within Arabia Terra, expressed by cones, knobs or large layered mounds. These mounds have lot in common with terrestrial mud volcanic morphologies, that present well-developed edifices, summit depressions and flow-like features, however earth-like mud volcanism require gaseous phases contributes which up to date can not be unambiguously detected on the Mars; this is indeed one of the primary objectives of ExoMars – TGO mission

The possibility to apply such an approach to other areas of the Martian surface, such as the widespread mud volcanic fields of Acidalia Planitia could help reconstruct the history of Mars' hydrologic cycle, subsurface water activity and fluid expulsion events both at a regional and local scale. Moreover, this approach will be useful for targeting possible fluid reservoir within large craters (Oehler and Etiope, 2017) and for further exploration in the framework of ExoMars TGO and targeting of CaSSIS imager observations (Thomas et al., 2017).

**Acknowledgements**

We gratefully thank the Editor, Dr. Jeffrey R. Johnson the two anonymous Referees who provided helpful and detailed suggestions for substantial manuscript improvement. We also thank Prof. Vikram Unnithan for careful review of the document and its English form.

This research was supported by ASI (Italian Space Agency) research grant No. I/060/10/0, the ASI-INAF agreement no.2017-03-17 and European Union's Horizon 2020 research and innovation program under grant agreement No 776276-PLANMAP.

**Appendix A. Testing the methods**

*A1. Mounds characterization and automatic extraction parameters*

To morphologically characterize mounds we have firstly applied the TPI combined with profile curvatures on the dataset from Pondrelli et al., (2011) resampling a higher resolution (1m) HiRISE DTM at CTX DTM resolution (18m ). This was done to calibrate our method in order to automatically detect the same features (filtering outliers) on other terrains in which only CTX DTMs where available.

Fig. A1 shows the profile curvatures belonging to specific geomorphological classes extracted with TPI from the DTM in the dataset mapped by Pondrelli et al., (2011). We have seen that the emergence of small hills and convex morphologies fall within both category 8 and 9 and thus we contoured these objects following the points of zero curvature on their flanks (the flex point). This is the only way to preserve the actual aspect ratio of these objects and avoid the interference of the rough sloping terrain at their base.

From an analysis on the aspect ratio in plan-view of the objects mapped in Pondrelli et al., (2011), we have seen that every mound has an aspect ratio >0.5. To get rid of ridges, artifacts and yardangs (that are often drop-shaped and elongated) we calculated the minimum bounding geometry of the contoured shapes, extracting major and minor axes, and filtering them according to their aspect ratios (i.e. every object with aspect ratio <0.5 was deleted).

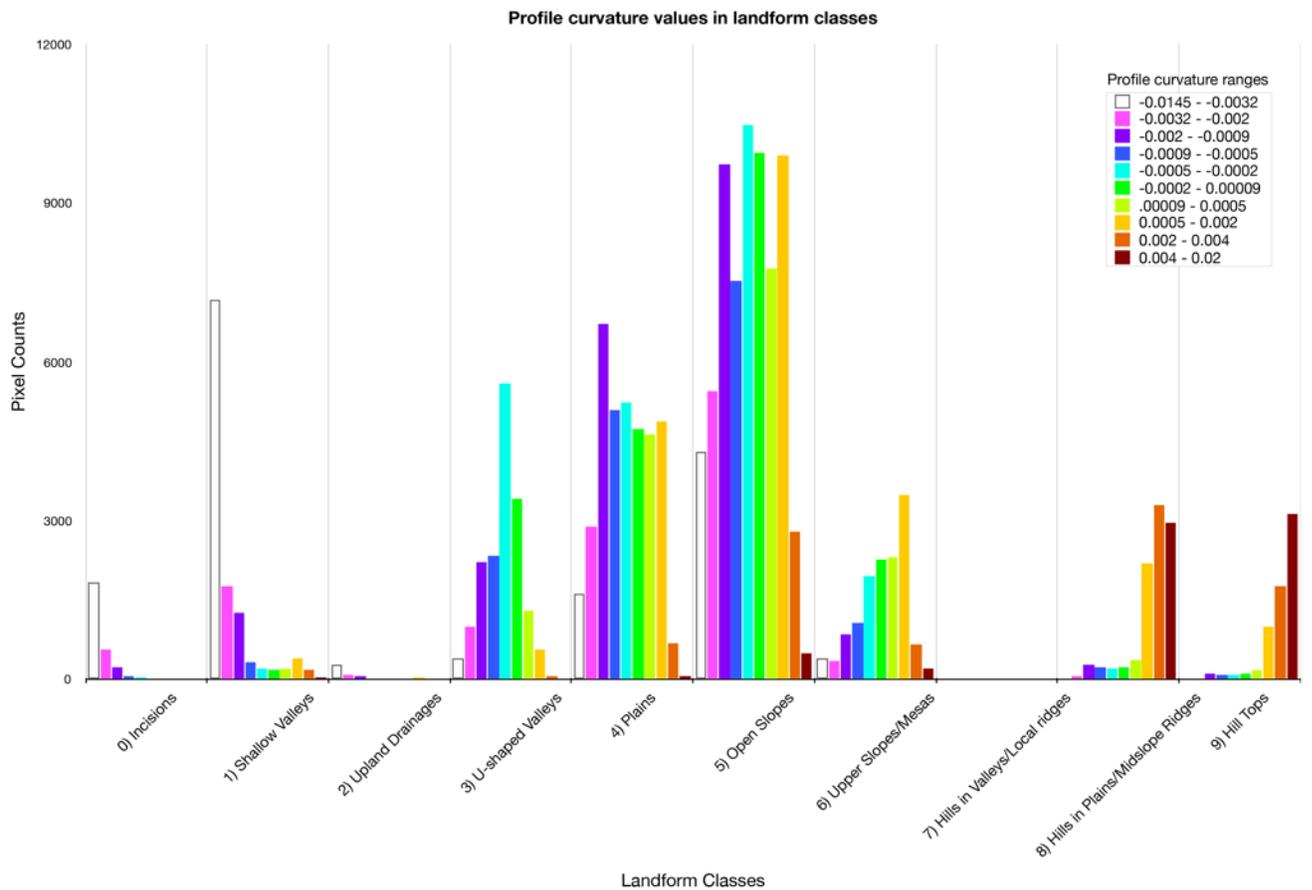

**Fig. A1**: Curvature ranges in each landform class defined by Jenness (2006) and Weiss (2001). It is clearly visible that, as expected, negative curvatures that highlight concavities fall within the incisions and shallow valleys identified in this case by the concavities at the base of the mounds and within them. The widest range of curvatures with a predominance between -0.002 and 0.002 fall in the landform classes that identify open slopes, broad plains, and flat-topped mesas. The highest curvatures >0.002 identify the positive reliefs and hilltops and are dominant in class 8 and 9. However, a distinction should be noted in these latter two classes being the frequency of the highest curvatures, ranging from 0.004 to 0.02, prominent in class 9, and comparable to 0.002-0.004 curvatures in class 8. In conclusion, all the putative mud volcanoes identified by Pondrelli et al., (2011) fall within class 8, whereas yardangs are always with sharp tops hence falling within features of high curvature in class 9.

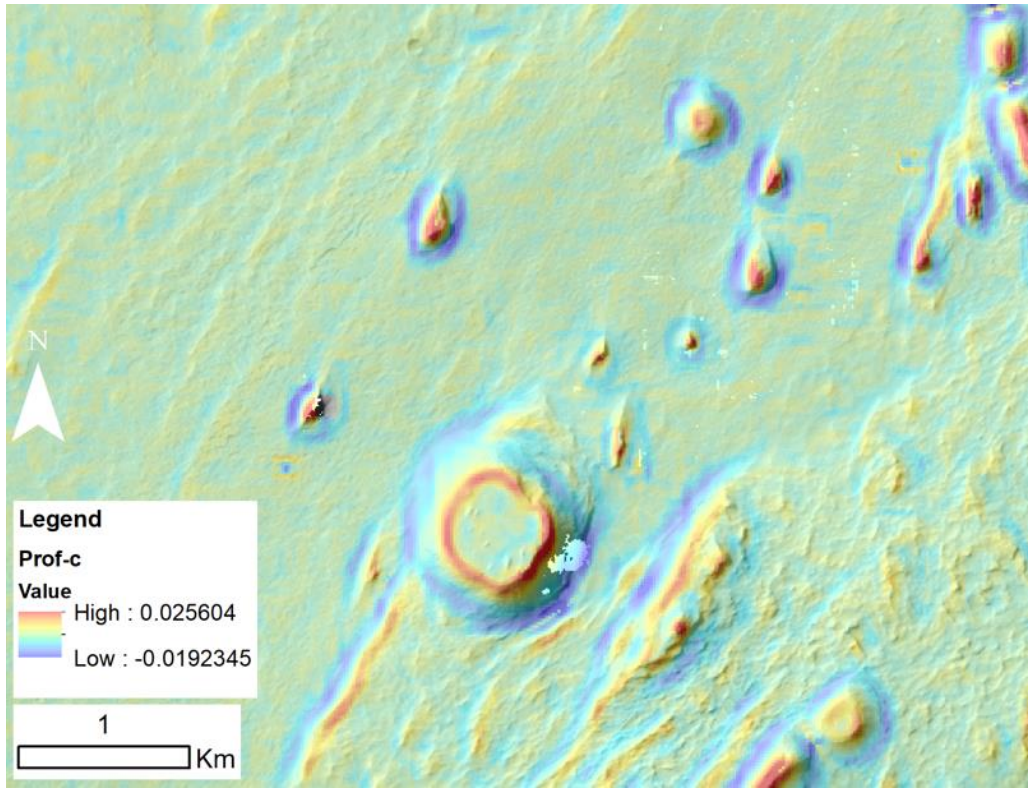

**Fig. A2**: Mesas are highlighted by the highest curvature values surrounding a zero-curvature area, being the flat top.

| Filtering Parameters | Values |
|:---:|:---:|
| Max Length | 300 m |
| Min Length | 50 m |
| Aspect ratio | >0.5 |
| Profile curvature | <0.004 |

**Table A1**: Filtering parameters used for automated mounds extraction on the south-eastern Firsoff subset.

However, several yardangs still remained in our automated mapping, but the analysis of profile curvatures showed that almost all of them present a specific set of curvatures mostly in category 9. In fact, both categories 8 and 9 present a prevalence of three ranges of curvatures: 0.0005-0.002, 0.002-0.004, 0.004-0.02 that are however differently distributed within the two categories. In category 8 the ranges or curvatures are almost equal, meaning that the related objects present gentle transition from low to high curvatures and results in a rounded profile. By contrast in category 9 the prominence of the sharpest curvatures (>0.004) result in a sudden transition form no curvature towards very sharp edges. This is typical of yardangs that, having an aspect ratio >0.5, exhibit a very sharp crest. Hence, we verified that the dataset from Pondrelli et al., 2011 mostly corresponds to category 8 and partly 9, with all the mounds presenting curvatures <0.004. It is still possible that some yardangs could actually be heavily eroded mounds, but we chose not to incorporate them in the analyses due to this uncertainty. Flat top mesas belong mostly to category 6, and have a sharp contact between 0.00008 and 0.001 curvatures. Visually, they are easily identifiable because they present an annular high-curvature region surrounding an almost flat portion (Fig. A2). By filtering them according to what exposed above and the parameters in table A1 ,we obtained an almost perfect match between the automatic extraction and the manual mapping (see Fig. 4). In addition to that, a final supervised inspection was performed on the obtained contouring shapefiles to locate possible outliers and ambiguities. To test the reliability of the automatic extraction method in correctly locating the mounds, we have compared the position of the centroids from the contoured features with those from Pondrelli et al., (2011). In Fig. A3 we visually represented the location analysis where we have calculated the minimum distance between automated and manually mapped points and evaluated their difference. In the plot in Fig. A4 it is visible that ~60% of the automatically extracted dataset fall within less than 20 meters from the points derived from Pondrelli et al. 2011 mapping, with another 10% falling within 60 m. The remaining 30% either belong to newly mapped points, that were not considered in the manual mapping, or to the subdivision of coalescent features into multiple features. Indeed, as it is visible in the subsets of Fig. A3, there are several cases of coalescing mounds, mapped as single morphologies in Pondrelli et a., (2011) that are actually composed of 4/5 objects almost equally distributed around the centroid of the composite mound as correctly detected by our algorithm. In fact, with the automatic extraction we were able to map 343 mounds versus the 259 of the manual mapping by Pondrelli et al. 2011. However, 44 out of 84 newly mapped features belong to 16 composite objects, whereas 40 features are effectively new detections. Hence, 70% of automatically mapped features are within less than 20 m from the manual mapped centroids, 12.8% correspond to coalescing mounds and 11.6% are newly mapped features that, from a visual analysis we are confident to asses that are actually mounds. The good reliability of our detection method is further confirmed by the consistency between the nearest neighbor analysis of the two datasets (table A2).

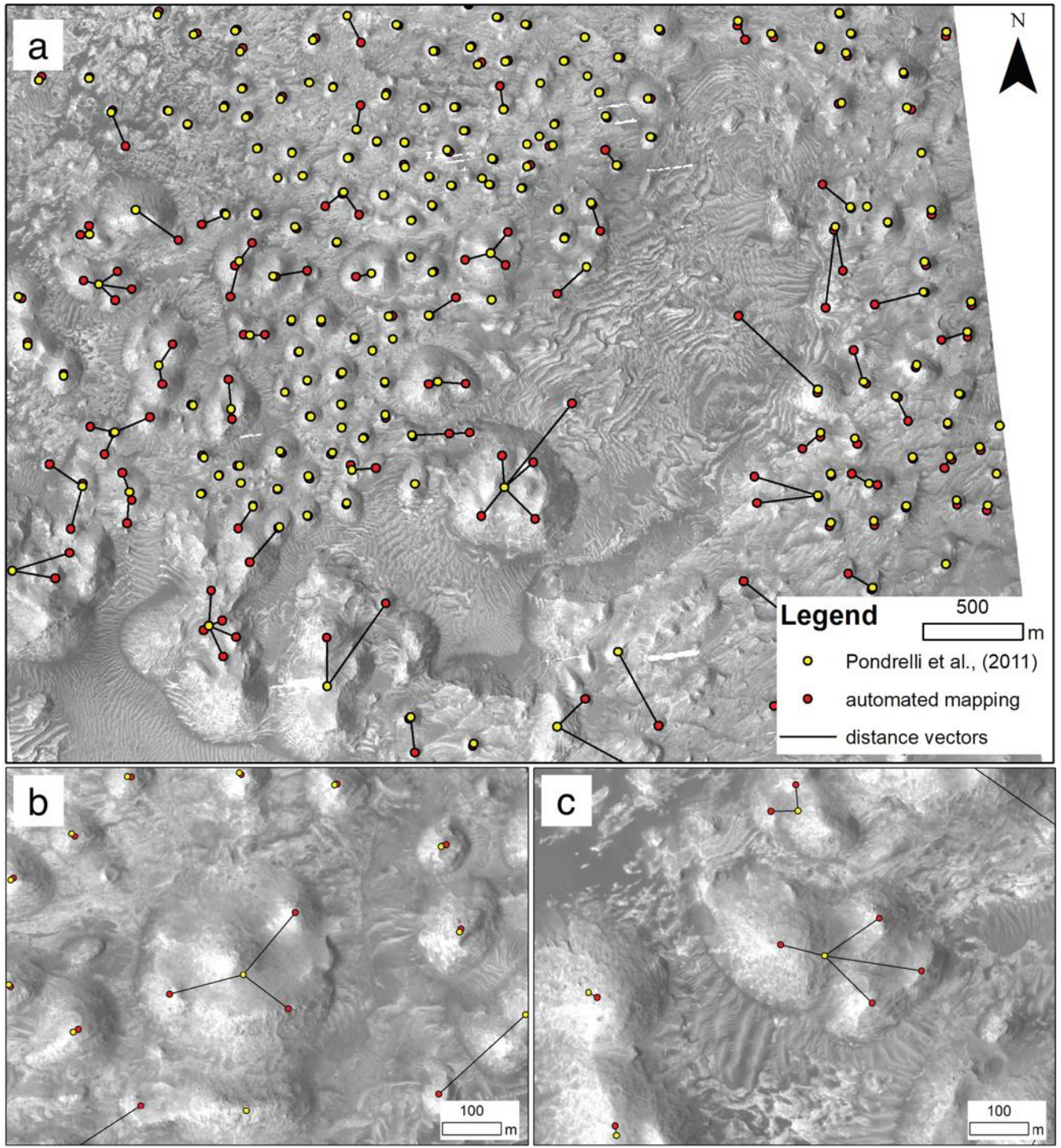

**Fig. A3**: Visual representation of the automated mapped mounds in red and their correspondence to the manually mapped ones, calculated as the minimum distance. Subsets b and c, show how composite edifices mapped as single features in Pondrelli et al., (2011) in reality were extracted as multiple mounds with TPI/profile curvature method.

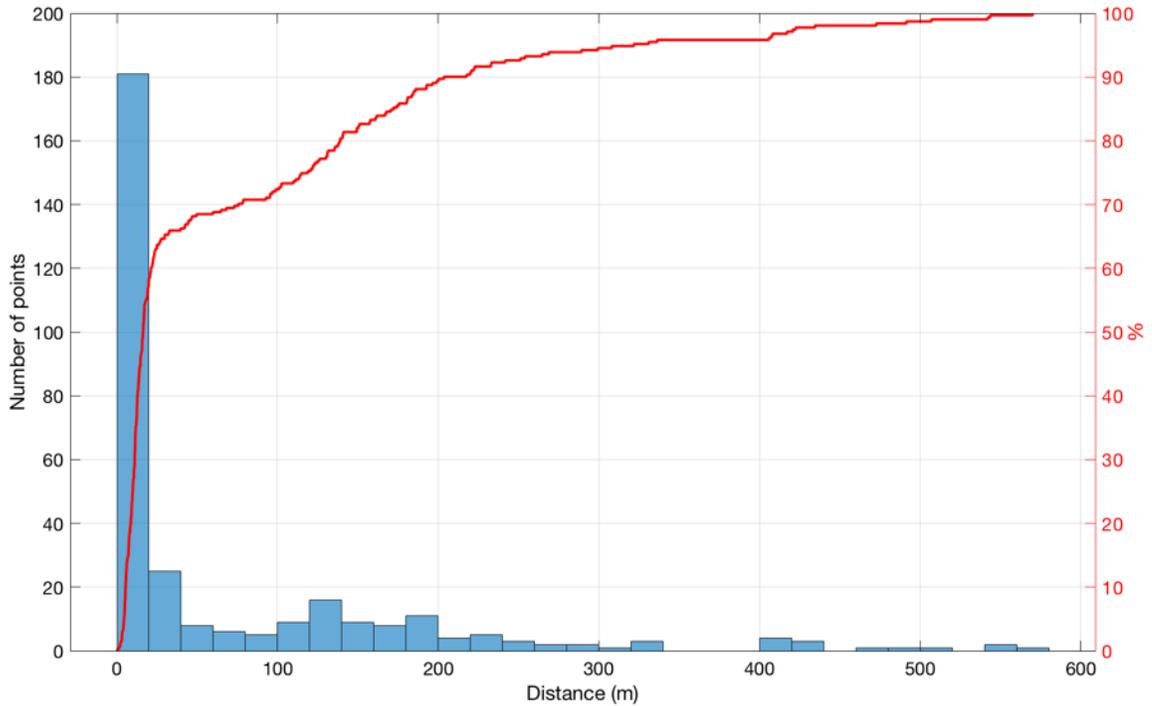

**Fig. A4**: Plot showing how the automatically extracted mounds are very proximal to the manually mapped mounds by Pondrelli et al., (2011). The left y axis presents the amount of points falling within the bins, whose width is set as 20 meters. The right axis shows the cumulative of the mounds in % according to their proximity. ~60% of the population falls within 20 meters (or less) while ~70% of the entire population falls within 60 meters. The remaining points are those related to coalescent mounds or newly mapped mounds (i.e., those >200 m of distance).

**Appendix B. Cutoff estimation for mounds spatial distribution**

Each point corresponding to mounds position has been analyzed according to the equations in (1,) and (2) resulting in the plots *l* vs *C(l)* that show how is the D average value. Additionally, the black box in the plots $R^2$ vs $\Delta log(l)$ in Fig. B1(a, b, c, d) contains the maximum fractal correlation (i.e. the size range between Lco and Uco) that presents the highest possible $R^2$ value for the largest possible size-range (*Lco-Uco*). The comparison of $R^2$ vs $\Delta log(l)$ with the local slope of *l* vs *C(l)* is shown for each analyzed dataset (fig B1).

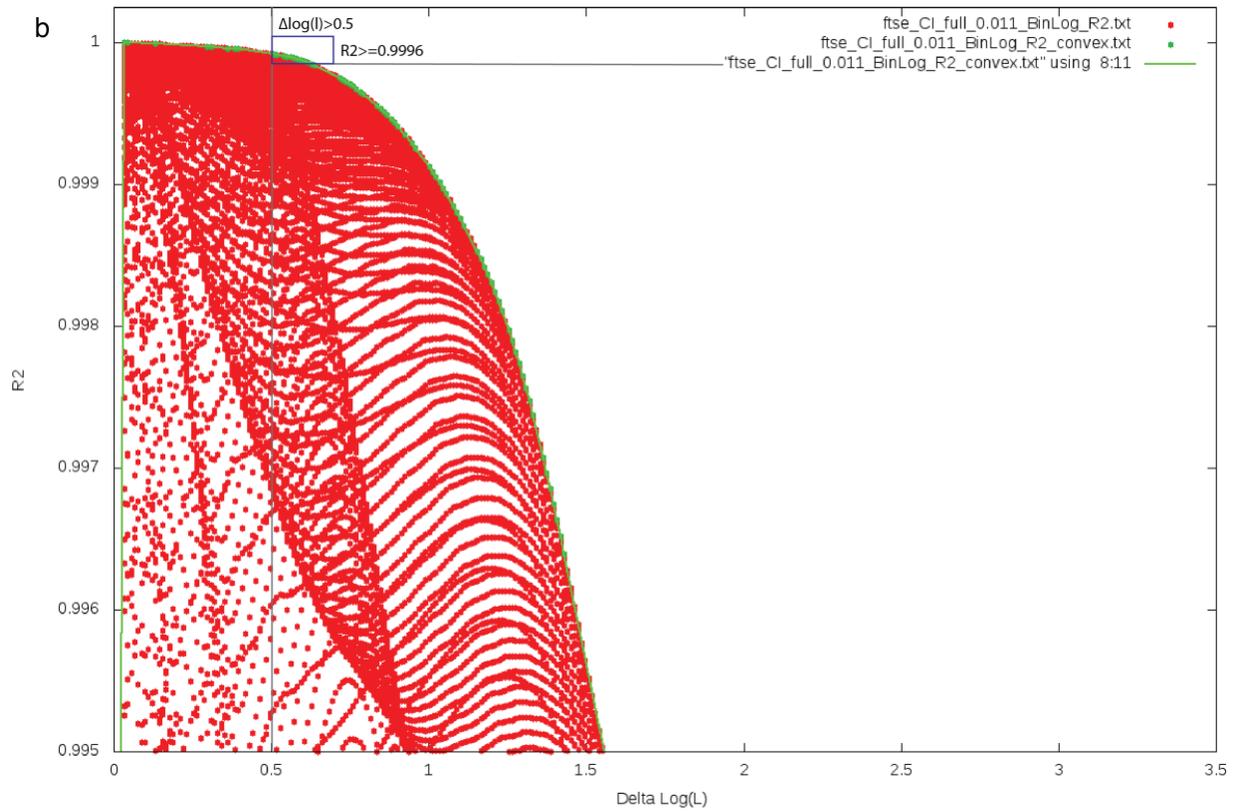

a
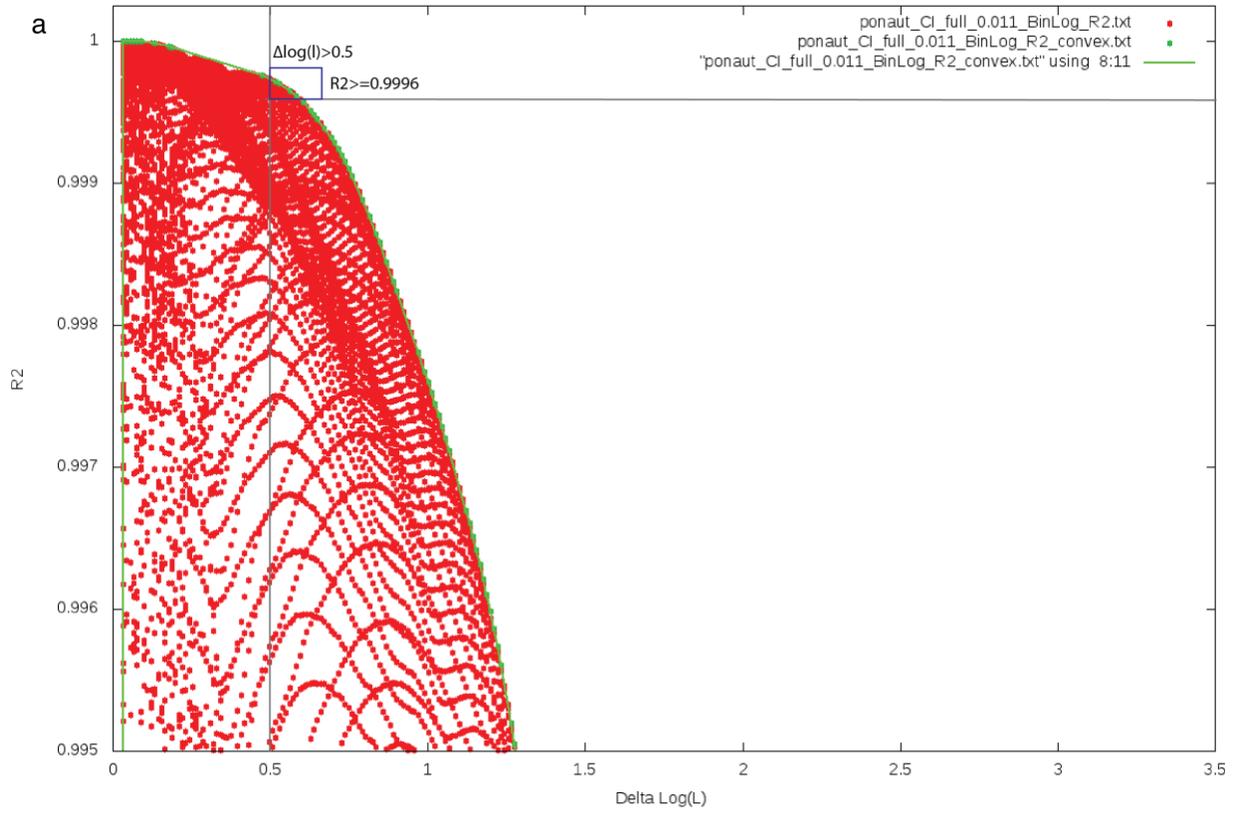

c
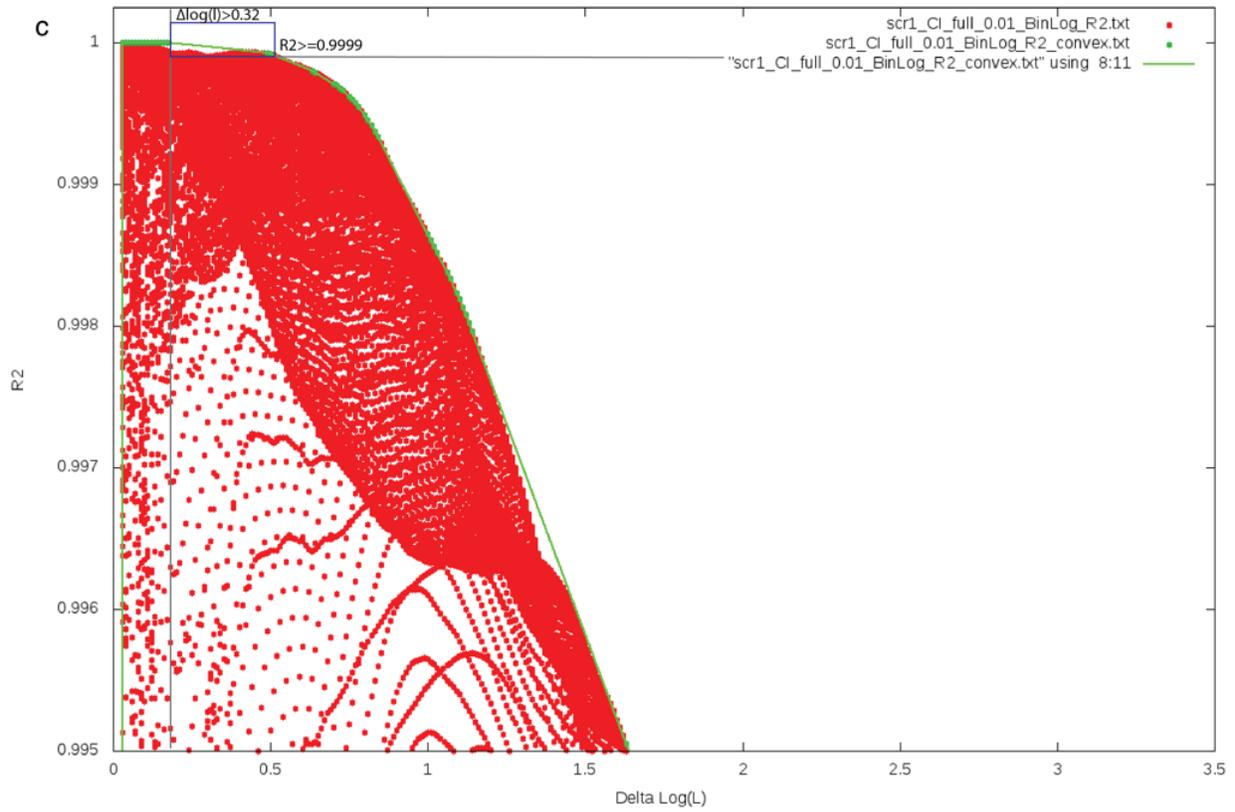

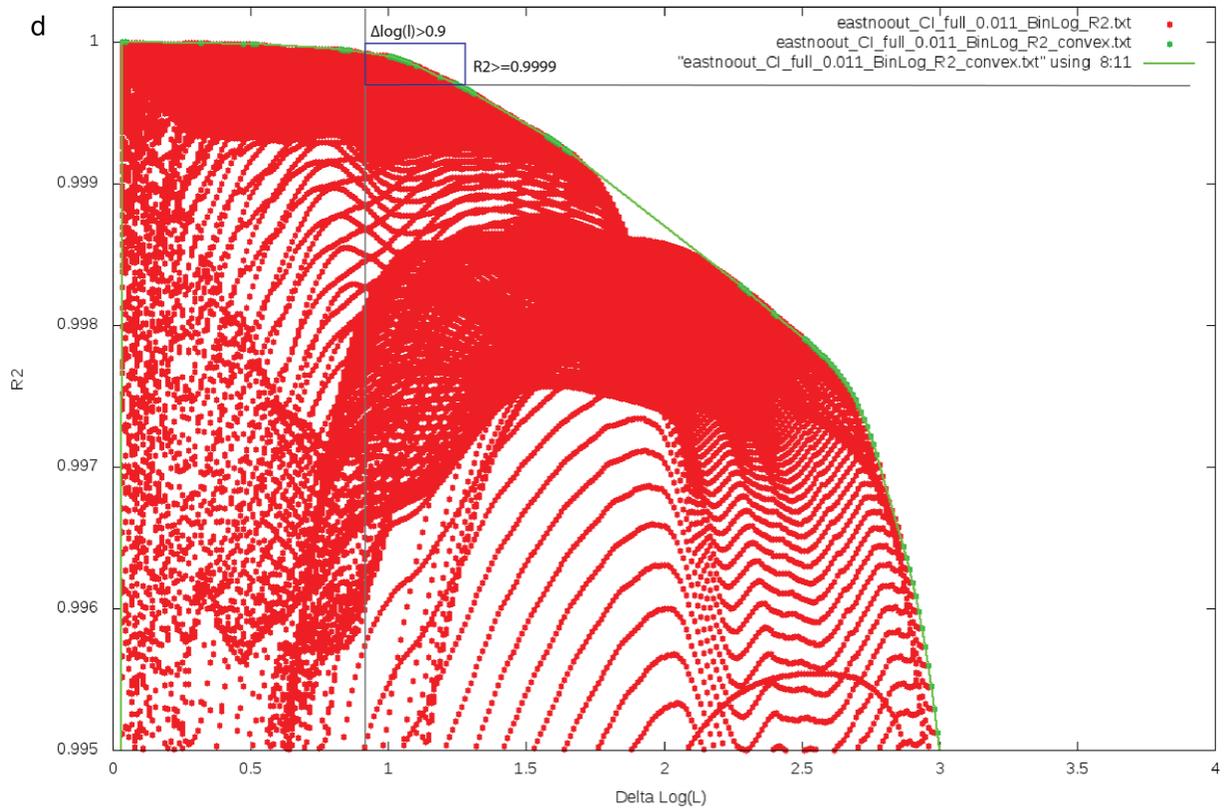

**Fig B1**: Plots showing all the possible $R^2$ vs $\Delta log(l)$ of all the population of the three craters, with the green line marking the border of the distribution. The grey lines represent the lower boundaries of the space containing the green line (the blue box) where the conditions in locating the Lco and Uco are satisfied (see Lucchetti et all., 2017 for further details). a) Firsoff subset used for testing the method, b) south-eastern Firsoff, c) Kotido crater, d) the unnamed eastern crater.

Mazzarini, F., Keir, D., Isola, I., 2013. Spatial relationship between earthquakes and volcanic vents in the central-northern Main Ethiopian Rift. J. Volcanol. Geotherm. Res. 262, 123–133. https://doi.org/10.1016/j.jvolgeores.2013.05.007

Mazzini, A., Etiope, G., 2017. Mud volcanism: An updated review. Earth-Science Rev. 168, 81–112. https://doi.org/10.1016/j.earscirev.2017.03.001

Melosh, H. J. Hydrocode equation of state for SiO2. Meteoritics & Planetary Science 42, 2035–2182 (2007).

Michalski, J.R., Bleacher, J.E., 2013. Supervolcanoes within an ancient volcanic province in Arabia Terra, Mars. Nature 502, 47–52. https://doi.org/10.1038/nature12482

Michalski, J.R., Cuadros, J., Niles, P.B., Parnell, J., Deanne Rogers, a., Wright, S.P., 2013. Groundwater activity on Mars and implications for a deep biosphere. Nat. Geosci. 6, 133–138. https://doi.org/10.1038/ngeo1706

Mumma, Michael J, Villanueva, G.L., Novak, R.E., Hewagama, T., Bonev, B.P., DiSanti, M.A., Mandell, A.M., Smith, M.D., 2009. Strong Release of Methane. Science (80-. ). 323, 1041–1045. https://doi.org/10.1126/science.1165243

Oehler, D. Z., & Allen, C. C. (2010). Evidence for pervasive mud volcanism in Acidalia Planitia, Mars. *Icarus*, *208*(2), 636–657. https://doi.org/10.1016/j.icarus.2010.03.031

Oehler, D. Z., & Etiope, G. (2017). Methane Seepage on Mars: Where to Look and Why. *Astrobiology*, *17*(xx), ast.2017.1657. https://doi.org/10.1089/ast.2017.1657

Orbach, R., 1986, Dynamics of fractal networks: Science, v. 231, p. 814–819, doi: 10.1126/science.231.4740.814.
50